\documentclass[prd,showpacs,letterpaper,10pt,aps,notitlepage,nofootinbib,superscriptaddress]{revtex4-2}
\usepackage[utf8]{inputenc}
\usepackage{xcolor}

\usepackage{physics}
\usepackage{slashed}
\usepackage{amsmath}
\usepackage{mathtools}
\usepackage{placeins}
\usepackage{amsfonts}
\usepackage{amssymb}

\usepackage{hyperref}
\hypersetup{colorlinks, linkcolor = [rgb]{0,0.0,0.75}, citecolor = [rgb]{0,0.0,0.75}, urlcolor = [rgb]{0,0.0,0.75}}

\date{September 18, 2025} 

\begin{document}

\title{Efficient lattice QCD computation of radiative-leptonic-decay form factors at multiple positive and negative photon virtualities}

\author{Davide Giusti}
\affiliation{Jülich Supercomputing Centre, Forschungszentrum Jülich, 52428, Jülich, Germany}
\affiliation{Fakultät für Physik, Universität Regensburg, 93040, Regensburg, Germany}

\author{Christopher F. Kane}
\affiliation{Maryland Center for Fundamental Physics and Department of Physics, University of Maryland, College Park, MD 20742, USA}
\affiliation{Joint Center for Quantum Information and Computer Science, University of Maryland, College Park,
Maryland 20742, USA.}

\author{Christoph Lehner}
\affiliation{Fakultät für Physik, Universität Regensburg, 93040, Regensburg, Germany}

\author{Stefan Meinel}
\affiliation{Department of Physics, University of Arizona, Tucson, AZ 85721, USA}

\author{Amarjit Soni}
\affiliation{Brookhaven National Laboratory, Upton, NY 11973, USA}

\begin{abstract}
In previous work [\href{https://doi.org/10.1103/PhysRevD.107.074507}{Phys. Rev. D 107, 074507 (2023)}], we showed that form factors for radiative leptonic decays of pseudoscalar mesons can be determined efficiently and with high precision from lattice QCD using the ``3d method,'' in which three-point functions are computed for all values of the current-insertion time and the time integral is performed at the data-analysis stage. Here, we demonstrate another benefit of the 3d method: the form factors can be extracted for any number of nonzero photon virtualites from the same three-point functions at no extra cost. We present results for the $D_s\to\ell\nu\gamma^*$ vector form factor as a function of photon energy and photon virtuality, for both positive and negative virtuality, for a single ensemble with 340 MeV pion mass and $0.11$ fm lattice spacing. 
In our analysis, we separately consider the two different time orderings and the different quark flavors in the electromagnetic current. 
We discuss in detail the behavior of the unwanted exponentials contributing to the three-point functions, as well as the choice of fit models and fit ranges used to remove them for various values of the virtuality.
While positive photon virtuality is relevant for decays to multiple charged leptons, negative photon virtuality suppresses soft contributions and is of interest in QCD-factorization studies of the form factors.
\end{abstract}

\maketitle

\section{Introduction}

In this paper, we present an extension of our previous study~\cite{Giusti:2023pot} on lattice-QCD methods for computing the hadronic matrix elements describing the weak decays of pseudoscalar mesons, $H$, into a lepton, a neutrino, and a photon, $\ell^+ \nu_\ell \gamma$ (or $\ell^+\ell^-\gamma$) by focusing on the more general case of an emitted off-shell photon, $\gamma^\ast$, \textit{i.e.}, $H^+ \to  \ell^+ \nu_\ell \gamma^\ast$ or $H^0 \to \ell^+ \ell^- \gamma^\ast$.

In Ref.~\cite{Giusti:2023pot} we have introduced and explored in detail a number of strategies to efficiently calculate with high accuracy in lattice QCD the structure-dependent axial-vector and vector form factors that, in the Standard Model, enter the radiative leptonic decay rates of pseudoscalar mesons at leading order in $\alpha_{\rm em}$. Prominent among these are the so-called ``3d method'' that helps with controlling the systematic uncertainties from unwanted exponentials in the sum over intermediate states in the extraction of the matrix elements contributing to the decay amplitudes, and the combined use of an infinite-volume approximation technique to investigate the full kinematically allowed photon-energy range of the form factors.
Precise determinations of the hadronic contributions to the radiative leptonic decay rates using lattice gauge theory can overcome the various limitations of approaches traditionally adopted in the literature, such as quark models \cite{Atwood:1994za,Colangelo:1996ct,Chang:1997re,Geng:2000fs,Chelkov:2001qx,Hwang:2005uk,Barik:2008zza,Shen:2013oua,Kozachuk:2017mdk,Dubnicka:2018gqg,Guadagnoli:2023zym,Guadagnoli:2023ddc}, QCD factorization/soft-collinear effective theory/perturbative QCD \cite{Korchemsky:1999qb,Beneke:1999br,Descotes-Genon:2002crx,Lunghi:2002ju,Braun:2012kp,Wang:2016qii,Beneke:2018wjp,Wang:2018wfj,Shen:2018abs,Beneke:2020fot,Shen:2020hsp,Wang:2021yrr,Galda:2022dhp,Cui:2023yuq}, light-cone sum rules \cite{Khodjamirian:1995uc,Ali:1995uy,Eilam:1995zv,Aliev:1996ud,Ball:2003fq,Janowski:2021yvz}, chiral perturbation theory \cite{Bijnens:1996wm,Geng:2003mt,Mateu:2007tr,Unterdorfer:2008zz,Cirigliano:2011ny,Burdman:1994ip}, and dispersion relations \cite{SalehKhan:2004kj,Guadagnoli:2017quo,Kurten:2022zuy}, being either model-dependent, making truncations in the relevant expansions, or requiring a large number of external inputs.
In addition to our previous lattice calculation \cite{Giusti:2023pot}, we also refer to the works of the Rome/Soton group~\cite{Desiderio:2020oej,Frezzotti:2020bfa,Frezzotti:2023ygt,Frezzotti:2024kqk,DiPalma:2025iud}.

Here we show that the 3d method allows to efficiently extract the form factors for any number of nonzero photon virtualities from the large-Euclidean-time limit of the same Euclidean three-point function of the real photon emission processes at no extra cost.
The analysis of correlation functions for positive photon virtualities allows to study rare processes of decays in three or four charged leptons,  i.e.~$H^+ \to \ell^{\prime +} \ell^{\prime -} \ell^+ \nu_\ell$ or $H^0 \to \ell^{\prime +} \ell^{\prime -} \ell^+ \ell^-$. These decays are sensitive probes of physics beyond the Standard Model; see {\it e.g.}, Refs.~\cite{Cirigliano:2011ny,Chala:2019vzu,Beneke:2021rjf}. Early lattice studies in the kaon sector have been carried out in Refs.~\cite{Tuo:2021ewr,Gagliardi:2022szw}, which used heavier-than-physical quark masses to avoid difficulties attributable to the presence of intermediate states that hinder analytic continuation from Euclidean to Minkowski spacetime. In this regard, recent lattice techniques of spectral-density reconstruction \cite{Hansen:2019idp,Frezzotti:2023nun} could be applied to overcome such difficulties, and the combined use of our 3d method can help achieve results at the physical point, alleviating signal-to-noise problems at large photon energies \cite{Frezzotti:2023ygt,DiPalma:2025iud}.
Interestingly, in contrast, the study of form factors for negative photon virtualities is entirely free from limitations due to the use of Euclidean correlation functions, as explained in Section~\ref{sec:Tmunu_C3_def}, and provides important complementary information to calculations performed in the framework of light-cone sum rules and QCD factorization \cite{Khodjamirian:1997tk,Wang:2016qii,Beneke:2018wjp}. In fact, in these regimes, soft contributions to heavy-meson form factors, which are not well constrained in the light-cone expansion, are suppressed, and a lattice calculation at different photon-virtuality values could help control them.

In this study, we make use of the ``24I'' RBC/UKQCD lattice gauge-field ensemble with 2+1 flavors of domain-wall fermions and the Iwasaki gauge action \cite{RBC:2010qam}, with inverse lattice spacing $a^{-1} = 1.785 (5)$ GeV and pion mass $m_\pi = 340 (1)$ MeV \cite{RBC:2014ntl}. We consider the process $D_s^+ \to  \ell^+ \nu_\ell \gamma^\ast$, for which we provide, for the first time, model-independent determinations of the vector form factor as a function of photon energy and virtuality, for both positive and negative virtualities. This paper focuses on a detailed investigation of lattice data-generation and data-analysis methods. Computations at the physical pion mass and for mesons other than the $D_s$, extrapolations to the continuum limit, and phenomenological studies of the decay observables are left for future work.

The structure of the remainder of this paper is as follows. In Sec.~\ref{sec:Tmunu_C3_def} we review how to relate the Minkowski-space hadronic tensor to a Euclidean-space three-point correlation function. We describe the 3d method in Sec.~\ref{sec:seq_prop}. The details of the lattice gauge-field ensemble and the lattice actions and parameters are given in Sec.~\ref{sec:lattice_setup}. Section~\ref{sec:imp_estimators} illustrates a number of improvements for determining the relevant form factors. The fit methods used to remove unwanted exponentials from intermediate and excited states are described in Sec.~\ref{sec:fit_methods}. The final form factors results at positive and negative virtuality are presented in Sec.~\ref{sec:final_res}, and we conclude in Sec.~\ref{sec:conclusions}.

\section{Hadronic tensor from Euclidean correlation functions}
\label{sec:Tmunu_C3_def}

\subsection{Hadronic tensor}
The decay we consider is that of a pseudoscalar meson $H$, with quark flavors $\bar{q}_2$ and $q_1$, into a lepton $\ell$, neutrino $\nu$ and off-shell photon $\gamma^*$.
The initial-state pseudoscalar meson has mass $m_H$, momentum $\vec{p}_H$, and the off-shell photon has virtuality $p_\gamma^2 = E_\gamma^2 - \vec{p}_\gamma^2 \neq 0$.
Using the weak effective Hamiltonian and working to first order in perturbation theory in the electromagnetic interaction, the remaining QCD physics left to calculate is contained in the hadronic tensor $T_{\mu \nu}$.
The hadronic tensor receives contributions from both the axial-vector and vector weak currents.
As discussed in the Introduction, in this work we focus on the vector component $T_{\mu \nu}^V$, which is given by
\begin{equation}
    T^V_{\mu \nu}(p_H, p_\gamma) = -i \int d t_\text{em} \int d^3 x \ e^{i p_\gamma \vdot x} \bra{0} \textbf{T} \big( J^\text{em}_\mu(t_\text{em},\vec{x}) V_\nu(0) \big) \ket{H(\vec{p}_H)} = \epsilon_{\mu \nu \tau \rho} p_\gamma^\tau p_H^\rho \frac{F_V}{m_H},
\end{equation}
where $J^\text{em}_\mu = \sum_q Q_q \bar{q} \gamma_\mu q$ is the electromagnetic current and $V_\nu = \bar{q}_1 \gamma_\nu q_2$ is the vector current component of the weak current.
The vector component of the hadronic tensor can be written in terms of a single form factor $F_V$, and is a function of the photon energy in the rest frame of the pseudoscalar meson, $E_\gamma^{(0)} = p_H \cdot p_\gamma/m_H$, and the virtuality of the photon, $p_\gamma^2$.
It is convenient to study the form factor as a function of the dimensionless ratio $x_\gamma = \frac{E_\gamma^{(0)}}{m_H/2}$, which takes values $0 \leq x_\gamma \leq 1 - \frac{m_\ell^2}{m_H^2}$ for physically allowed kinematics.
Unless stated otherwise, we will refer to $T_{\mu \nu}^V$ simply as the hadronic tensor for brevity.

To relate the Minkowski-time hadronic tensor to a Euclidean correlation function, we will compare the spectral decompositions of the different time-orderings of the two currents $J_\mu(t_{\rm em}, \vec{x})$ and $V_\nu(0)$. 
Inserting a complete set of energy-momentum eigenstates and performing the integral over $t_{\rm em}$ gives~\cite{Kane:2019jtj, Kane:2021zee, Giusti:2023pot}
\begin{equation}
	\begin{split}
		T^{V,<}_{\mu \nu} &= -i \int_{-\infty(1-i\epsilon)}^0 dt_{\text{em}} \int d^3x \ e^{-i p_\gamma \vdot x} \bra{0} V_\nu(0) J^\text{em}_\mu(t_{\text{em}}, \vec{x}) \ket{H(\vec{p}_H)}
		\\
		&= -\sum_{n} \frac{\bra{0} V_\nu(0) \ket{n(\vec{p}_H-\vec{p}_\gamma)} \bra{n(\vec{p}_H-\vec{p}_\gamma)} J^\text{em}_\mu(0) \ket{H(\vec{p}_H)}}{2 E_{n,\vec{p}_H-\vec{p}_\gamma}(E_\gamma + E_{n,\vec{p}_H-\vec{p}_\gamma} - E_{H,\vec{p}_H} - i \epsilon)},
	\end{split}
	\label{eq:hadronic_tensor_spectral_neg}
\end{equation}
and
\begin{equation}
	\begin{split}
		T^{V,>}_{\mu \nu} &= -i \int_{0}^{\infty(1-i\epsilon)} dt_{\text{em}} \int d^3x \ e^{-i p_\gamma \vdot x} \bra{0} J^\text{em}_\mu(t_{\text{em}}, \vec{x}) V_\nu(0) \ket{H(\vec{p}_H)}
		\\
		&= \sum_{m} \frac{\bra{0} J^\text{em}_\mu(0) \ket{m(\vec{p}_\gamma)} \bra{m(\vec{p}_\gamma)} V_\nu(0) \ket{H(\vec{p}_H)}}{2 E_{m,\vec{p}_\gamma}(E_\gamma - E_{m,\vec{p}_\gamma} - i \epsilon)}.
	\end{split}
	\label{eq:hadronic_tensor_spectral_pos}
\end{equation}

\subsection{Euclidean three-point function with weak current at origin}
We use the Euclidean three-point correlation function
\begin{equation}
	C_{3, \mu \nu}^V(t_{\text{em}}, t_H) = \int d^3x \int d^3y \ e^{-i \vec{p}_\gamma \vdot \vec{x}} e^{i \vec{p}_H \vdot \vec{y}}  \langle J_{\mu}^\text{em}(t_{\text{em}}, \vec{x}) V_\nu(0) \phi^\dagger_H(t_H, \vec{y}) \rangle
	\label{eq:three_point}
\end{equation}
to calculate the hadronic tensor,
where $\phi_H^\dagger = -\bar{q}_2 \gamma_5 q_1$ is the meson interpolating field and $t_H$ is the source-sink separation.
We define time-integrated correlation functions for each time ordering as
\begin{equation}
    I_{\mu \nu}^{V,<}(t_H, T) = \int_{-T}^0 dt_{\rm em} \, e^{E_\gamma t_{\rm em}}C_{3, \mu \nu}^V(t_{\text{em}}, t_H), \qquad I_{\mu \nu}^{V,>}(t_H, T) = \int_{0}^T dt_{\rm em} \, e^{E_\gamma t_{\rm em}}C_{3, \mu \nu}^V(t_{\text{em}}, t_H),\,
\end{equation}
and their sum $I_{\mu \nu}^{V}(t_H, T) = I_{\mu \nu}^{V,<}(t_H, T)+I_{\mu \nu}^{V,>}(t_H, T)$.
By inserting two complete sets of energy-momentum eigenstates and performing the integral over Euclidean time $t_{\rm em}$, we find the following spectral decompositions:
\begin{align}
    \begin{split}\label{eq:Imunu_spectral_neg}
        I^{V,<}_{\mu \nu}(t_H, T) &= \sum_{l,n} \frac{\bra{0} J^\text{weak}_\nu(0) \ket{n(\vec{p}_H - \vec{p}_\gamma)} \bra{n(\vec{p}_H - \vec{p}_\gamma)} J^\text{em}_\mu(0) \ket{l(\vec{p}_H)} \bra{l(\vec{p}_H)} \phi^\dagger_H(0) \ket{0}}{2E_{n, \vec{p}_H - \vec{p}_\gamma} 2E_{l, \vec{p}_H}(E_\gamma + E_{n,\vec{p}_H - \vec{p}_\gamma} - E_{l,\vec{p}_H})}
       \\
       &\hspace{0.25in} \times e^{E_{l, \vec{p}_H} t_H}\Big[1 - e^{-(E_\gamma - E_{l, \vec{p}_H} + E_{n, \vec{p}_H-\vec{p}_\gamma})T}\Big],
    \end{split}
    \\
    \begin{split}\label{eq:Imunu_spectral_pos}
        I^{V,>}_{\mu \nu}(t_H, T) &= -\sum_{l,m} \frac{\bra{0} J^\text{em}_\mu(0) \ket{m(\vec{p}_\gamma)} \bra{m(\vec{p}_\gamma)}  J^\text{weak}_\nu(0)  \ket{l(\vec{p}_H)} \bra{l(\vec{p}_H)}  \phi^\dagger_H(0) \ket{0}}{2E_{m, \vec{p}_\gamma}2E_{l, \vec{p}_H}(E_\gamma - E_{m,\vec{p}_\gamma})} 
        \\
        &\hspace{0.25in} \times e^{E_{l,\vec{p}_H} t_H} \big[ 1-e^{(E_\gamma - E_{m, \vec{p}_\gamma})T} \big].
    \end{split}
\end{align}
Looking at Eqs.~\eqref{eq:Imunu_spectral_neg} and \eqref{eq:Imunu_spectral_pos}, we see that the ground-state contribution can be isolated by taking the source-sink separation $t_H$ to large negative values.

After isolating the ground-state contribution by taking the source-sink separation $t_H$ to large negative values, we observe that each term in the spectral decomposition of $I_{\mu \nu}^{V,>}$ (Eq.~\eqref{eq:Imunu_spectral_neg}) is equal to the associated term in $T_{\mu \nu}^{V, <}$ (Eq.~\eqref{eq:hadronic_tensor_spectral_neg}) up to $1 - e^{-(E_\gamma - E_{H,\vec{p}_H} + E_{n, \vec{p}_H-\vec{p}_\gamma})T}$ factors.
In a similar way, each term in the spectral decomposition of $I_{\mu \nu}^{V,>}$ (Eq.~\eqref{eq:Imunu_spectral_pos}) is equal to the associated term in $T_{\mu \nu}^{V, >}$ (Eq.~\eqref{eq:hadronic_tensor_spectral_pos}) up to $1 - e^{-(E_\gamma - E_{m, \vec{p}_\gamma})T}$ factors.
To ensure that one can analytically continue to Minkowski time, these unwanted exponentials must decay with increasing $T$, which implies that the following inequalities must hold :
\begin{align}
    &t_{\rm em}<0: \qquad E_\gamma-E_{H,\vec{p}_H} + E_{n,\vec{p}_H-\vec{p}_\gamma}>0 \label{eq:tem_neg_condition},
    \\
    &t_{\rm em}>0: \qquad E_\gamma-E_{m,\vec{p}_\gamma}<0 \label{eq:tem_pos_condition}\,.
\end{align}

\begin{figure}
    \centering
    \includegraphics[width=0.95\linewidth]{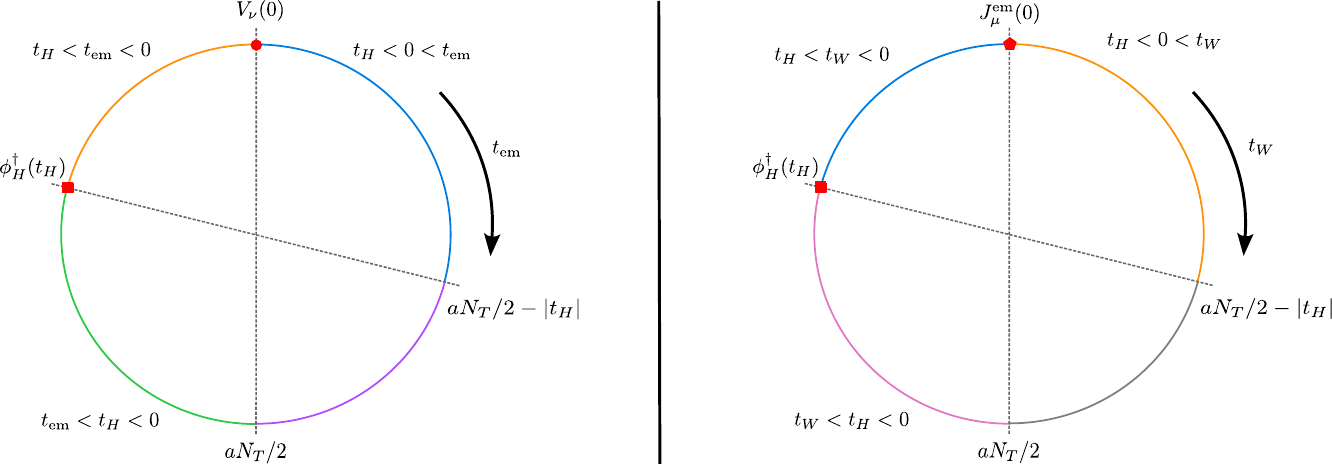}
    \caption{The left and right figures depict the different time orderings for the Euclidean three-point functions with the weak current (Eq.~\eqref{eq:three_point}) and EM current (Eq.~\eqref{eq:C3_EM}) at the origin, respectively.
    In the left (right) figure, the $t_{\rm em}$ ($t_W$) coordinate indicates the location of the EM (weak) current, with the clockwise direction corresponding to increasing time.
    Due to periodic boundary conditions, the time coordinate forms a closed circle. 
    For the left (right) figure, the weak (EM) current is fixed to time $t_{\rm em}=0$ ($t_W=0$).
    For both circles, the orange and blue arcs correspond to the $T^{V,<}_{\mu \nu}$ and $T^{V,>}_{\mu \nu}$ time orderings of the hadronic tensor.
    Importantly, the time orderings the orange and blue arcs correspond to in the left and right figures are flipped relative to each other; in the left figure, the orange and blue arcs correspond to the $t_{\rm}<0$ and $t_{\rm em}>0$ time orderings, while in the right figure the orange and blue arcs correspond to the $t_W>0$ and $t_W<0$ time orderings.
    Looking at the orange arcs, we see that, while for the $t_{\rm em}<0$ time ordering one is limited to values of integration range $T < |t_H|$, for the $t_W>0$ time ordering one has access to larger values of integration range $T < aN_T/2 - |t_H|$, where $N_T$ is the number of sites in the Euclidean time direction.
    In a similar way, looking at the blue arcs the roles flip; for the $t_W<0$ time ordering we are limited to $T < |t_H|$ and for the $t_{\rm em}>0$ time ordering we have reach $T < aN_T/2-|t_H|$.
    The purple, green, pink, and grey arcs correspond to unphysical time orderings.}
    \label{fig:time_orderings}
\end{figure}

Starting with the $t_{\rm em}<0$ time ordering, because the EM current cannot change quark-flavor quantum numbers, the intermediate states $\ket{n(\vec{p}_H-\vec{p}_\gamma)}$ must have the same quark flavor quantum numbers as the initial state pseudoscalar meson $H$.
The lowest-energy state with these quantum numbers is the pseudoscalar meson itself.
From this we see that the condition for $t_{\rm em}<0$ in Eq.~\eqref{eq:tem_neg_condition} is always satisfied except for $(E_\gamma, \vec{p}_\gamma) = (0, \vec{0})$\footnote{One can use parity symmetry to show that the pseudoscalar state $H$ does not contribute to the spectral decomposition of $I_{\mu \nu}^{V,<}$; the lowest-energy state that contributes is the associated vector meson $H^*$.
This implies that the analytic continuation from Euclidean to Minkowski time is safe even at $(E_\gamma, \vec{p}_\gamma) = (0, \vec{0})$.}, even for negative values of photon virtuality $p_\gamma^2<0$.
One important observation is that, while the unwanted exponentials do decay with $T$, for a given value of $t_H$, one only has access to $I_{\mu \nu}^{V,<}(t_H, T)$ for $0 < T < -t_H$.
The reason for this is that, as one increases $T$, one is integrating \emph{toward} the interpolating field, as depicted in Fig.~\ref{fig:time_orderings}.
We discuss the implications of the limited range of $T$ on our analysis in more detail in Sec.~\ref{sec:fit_methods}.

For the $t_{\rm em}>0$ time ordering, the intermediate states $\ket{m(\vec{p}_\gamma)}$ are flavorless, and the lowest-energy state that contributes is the two-pion state.
Rearranging Eq.~\eqref{eq:tem_pos_condition} gives the condition $p_\gamma^2 < 4m_\pi^2$.
Unlike the $t_{\rm em}<0$ case, there are theoretical limitations placed on the values of $p_\gamma^2$ that can be calculated using Euclidean lattice QCD.  In a finite volume,
these limitations can be reduced by separating out a discrete set of individual states but a detailed discussion of such an extension is left for future work.
One important observation, however, is that one can safely study negative virtuality $p_\gamma^2<0$ without having to worry about growing exponentials.
Unlike the $t_{\rm em}<0$ time ordering, increasing $T$ corresponds to integrating \emph{away} from the interpolating field; for a lattice with Euclidean time extent $aN_T$, the range of $T$ one can study $I_{\mu \nu}^{V,>}(t_H, T)$ is $0 < T < aN_T/2$, independent of $t_H$ (see Fig.~\ref{fig:time_orderings} for a visualization).

In general, we see that we can extract the different time orderings of the hadronic tensor by
\begin{align}
    T^{V,<}_{\mu \nu} &= - \lim_{T\to\infty}\lim_{t_H\to -\infty} \frac{2E_H e^{-E_H t_H}}{\bra{H(\vec{p}_H)} \phi^\dagger(0) \ket{0}} I_{\mu \nu}^{V,>}(t_H, T), \quad p_\gamma \neq (0,0,0,0),
    \\
    T^{V,>}_{\mu \nu} &= - \lim_{T\to\infty}\lim_{t_H\to -\infty} \frac{2E_H e^{-E_H t_H}}{\bra{H(\vec{p}_H)} \phi^\dagger(0) \ket{0}} I_{\mu \nu}^{V,>}(t_H, T), \quad p_\gamma^2 < 4 m_\pi^2\,.
\end{align}
We discuss our procedure for taking the $T \to \infty$ and $t_H \to -\infty$ limits in more detail in Sec.~\ref{sec:fit_methods}.
Throughout this work, we will often refer to $I_{\mu \nu}^{V,<}(t_H, T)$ and $I_{\mu \nu}^{V,>}(t_H, T)$ as simply the $t_{\rm em}<0$ and $t_{\rm em}>0$ datasets, respectively.

\subsection{Euclidean three-point function with electromagnetic current at origin}
Rather than use a three-point function with the weak current fixed to the origin, it was shown in Ref.~\cite{Giusti:2023pot} that translational invariance can be exploited to use an alternative three-point function with the EM current fixed to the origin instead.
In particular, $T_{\mu \nu}$ can also be extracted using
\begin{align}
    C_{3,\mu \nu}^{V, \text{EM}}(t_W, t_H) = e^{E_H t_W}\int d^3x \int d^3y \, e^{i(\vec{p}_\gamma-\vec{p}_H) \cdot \vec{x}} e^{i \vec{p}_H \cdot \vec{y}} \langle J_{\mu}^\text{em}(0) V_\nu(t_W, \vec{x}) \phi^\dagger_H(t_H, \vec{y}) \rangle\,.
    \label{eq:C3_EM}
\end{align}
where the superscript EM differentiates between the three-point function in Eq.~\eqref{eq:three_point} with the weak current at the origin.
Note that, relative to the three-point function in Eq.~\eqref{eq:three_point}, this correlation function has additional factors of $e^{E_H t_W}$ and $e^{-i \vec{p}_H \cdot \vec{x}}$.
Additionally, the phase factor $e^{i \vec{p}_\gamma \cdot \vec{x}}$ has a minus sign in the exponent relative to the factor $e^{-i \vec{p}_\gamma \cdot \vec{x}}$ in Eq.~\eqref{eq:three_point}.  
These modifications are required to shift the operators relative to each other, ensuring that the intermediate states in the time-integrated correlation function have both the correct momentum and come with the correct factors of energy in the denominator.

In a similar way, one can define time-integrated correlation functions 
\begin{align}
    &I^{V, \text{EM}, >}_{\mu \nu}(t_H, T) = \int_0^T dt_W \, e^{-E_\gamma t_W} C^{V, {\rm EM}}_{3,\mu \nu}(t_W, t_H), \quad I^{V, \text{EM}, <}_{\mu \nu}(t_H, T) = \int_{-T}^0 dt_W \, e^{-E_\gamma t_W} C^{V, {\rm EM}}_{3,\mu \nu}(t_W, t_H)\,,
\end{align}
where the superscript $>$ and $<$ now correspond to the $t_W>0$ and $t_W<0$ time orderings, respectively.
Inserting complete sets of energy-momentum eigenstates as before, their spectral decompositions are
\begin{align}
    I^{V, \text{EM}, >}_{\mu \nu}(t_H, T) &= \sum_{n,l} \frac{\bra{0} V_\nu(0) \ket{n(\vec{p}_H-\vec{p}_\gamma)} \bra{n(\vec{p}_H-\vec{p}_\gamma)} J_\mu(0) \ket{l(\vec{p}_H)} \bra{l(\vec{p}_H)} \phi^\dagger_H \ket{0}}{2E_{n,\vec{p}_H-\vec{p}_\gamma} 2E_{l, \vec{p}_H}(E_\gamma + E_{n,\vec{p}_\gamma - \vec{p}_H} - E_H)} \nonumber
    \\
    &\hspace{0.4in} \times e^{E_{l,\vec{p}_H} t_H} \Big[1- e^{-(E_\gamma + E_{n,\vec{p}_\gamma - \vec{p}_H} - E_H) T} \Big]
    \label{eq:Imunu_spectral_decomp_EM_greaterthan}
\end{align}
and
\begin{align}
    I^{V, \text{EM}, <}_{\mu \nu}(t_H, T) &= \sum_{m,l} \frac{\bra{0} J_\mu(0) \ket{m(\vec{p}_\gamma)} \bra{m(\vec{p}_\gamma)} V_\nu(0) \ket{l(\vec{p}_H)} \bra{l(\vec{p}_H)} \phi^\dagger_H \ket{0}}{2E_{m,\vec{p}_\gamma} 2E_{l,\vec{p}_H}(E_\gamma-E_{m,\vec{p}_\gamma} + \Delta E_{l, \vec{p}_H})} \nonumber
    \\
    &\hspace{0.4in} \times e^{E_{l,\vec{p}_H} t_H} \Big[ e^{(E_\gamma - E_{n,\vec{p}_\gamma} + \Delta E_{l, \vec{p}_H})T} - 1 \Big]\,,
    \label{eq:Imunu_spectral_decomp_EM_lessthan}
\end{align}
where we denote by $\Delta  E_l = E_{l,\vec{p}_H}-E_{H,\vec{p}_H}$ the excited-state energy gap for the $l^\text{th}$ excited state created by the interpolating field; this parameter is zero when $E_{l}$ corresponds to the ground state.
As expected, these expressions are similar to those in Eq.~\eqref{eq:hadronic_tensor_spectral_neg} and Eq.~\eqref{eq:hadronic_tensor_spectral_pos} except for the fact that the $t_W>0$ and $t_W<0$ time orderings correspond to the $t_{\rm em}<0$ and $t_{\rm em}>0$ time orderings, respectively.
Visualized in Fig.~\ref{fig:time_orderings}, for a given $t_H$, one has access to $I^{V,<,{\rm EM}}_{\mu \nu}(t_H, T)$ for $0 < T < -t_H$ and $I^{V,>,{\rm EM}}_{\mu \nu}(t_H, T)$ for $0 < T < aN_T/2$.
This fact has important implications for the behavior of the unwanted exponentials as a function of $t_H$ and $T$, which we discuss in more detail in Sec.~\ref{sec:fit_methods}.

For large source-sink separation $t_H$, one can use similar arguments as in the previous section to show that the unwanted exponentials that come with each intermediate state go to zero for large $T$ as long as 
\begin{align}
    &t_W>0: \qquad E_\gamma-E_{H,\vec{p}_H} + E_{n,\vec{p}_H-\vec{p}_\gamma}>0 \label{eq:tem_neg_condition_W},
    \\
    &t_W<0: \qquad E_\gamma-E_{m,\vec{p}_\gamma}<0 \label{eq:tem_pos_condition_W}\,,
\end{align}
which implies that the hadronic tensor can be calculated as
\begin{align}
    T^{V,<}_{\mu \nu} &= - \lim_{T \to \infty} \lim_{t_H \to -\infty} \frac{2 E_{H, \vec{p}_H} e^{-E_{H, \vec{p}_H} t_H}}{\bra{H(\vec{p}_H)} \phi^\dagger_H(0) \ket{0}} I^{\text{EM}, V, >}_{\mu \nu}(t_H, T), \quad p_\gamma \neq (0,0,0,0),
    \\
    T^{V,>}_{\mu \nu} &= - \lim_{T \to \infty} \lim_{t_H \to -\infty} \frac{2 E_{H, \vec{p}_H} e^{-E_{H, \vec{p}_H} t_H}}{\bra{H(\vec{p}_H)} \phi^\dagger_H(0) \ket{0}} I^{\text{EM}, V, <}_{\mu \nu}(t_H, T)\,, \quad p_\gamma^2 < 4 m_\pi^2\,.
\end{align}
Throughout this work, we will often refer to $I_{\mu \nu}^{V,<, {\rm EM}}(t_H, T)$ and $I_{\mu \nu}^{V,>, {\rm EM}}(t_H, T)$ as simply the $t_W<0$ and $t_W>0$ datasets, respectively.

\section{All photon virtualities for free using a 3d sequential propagator}
\label{sec:seq_prop}

As discussed in detail in Ref.~\cite{Giusti:2023pot}, there are two general methods for calculating the time-integrated correlation function $I_{\mu \nu}(t_H, T)$ on the lattice, referred to as the 3d and 4d methods.
In this section, we compare the number of propagator inversions required to calculate $I_{\mu \nu}(t_H, T)$ using the 3d and 4d methods, and show that using the 3d method, one calculates all photon virtualities for free.
The total cost of each method is summarized in Table~\ref{tab:num_solves}.
For a detailed discussion comparing more aspects of each method, we refer the reader to Ref.~\cite{Giusti:2023pot}.

\begin{figure}
    \centering
    \includegraphics[width=0.97\linewidth]{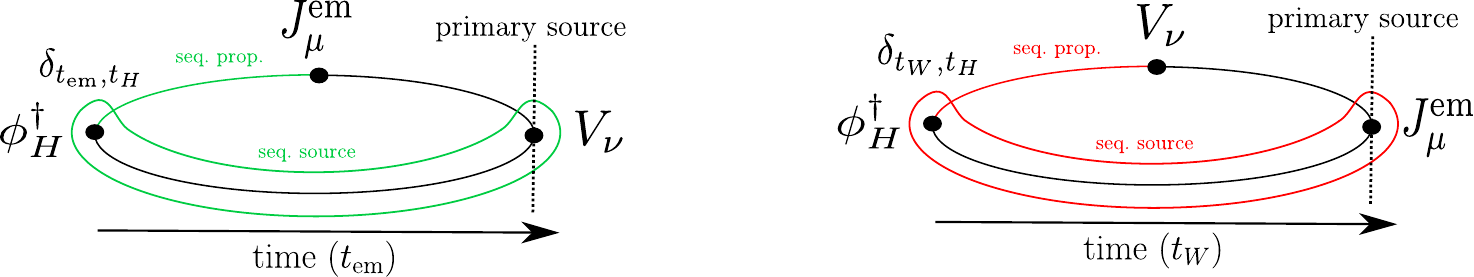}
    \caption{The left and right figures show schematic visualizations of the 3d method for the three point function with the weak and EM currents fixed to the origin, respectively. In the left (right) figure, the initial source is placed at the weak (EM) current location, the sequential source is circled in green (red) and the sequential propagator is shown with a green (red) line.}
    \label{fig:3d_method}
\end{figure}

The 4d method uses a four-dimensional sequential propagator through the electromagnetic current.
Using the 4d method, one calculates the time-integrated correlation function $I_{\mu \nu}(t_H, T)$ directly on the lattice; for a single value of $T$, one calculates all values of $t_H$ for free.
Because the phase factor $e^{-i \vec{p}_\gamma \cdot \vec{x}}$ and the exponential factor $e^{E_\gamma t_{\rm em}}$ (and also the $\gamma_\nu$ matrix associated with the weak current) must be included in the source for the sequential solve, calculating $I_{\mu \nu}(t_H, T)$ using the 4d method requires additional solves for each choice of photon 4-momentum $p_\gamma = (E_\gamma, \vec{p}_\gamma)$, and integration range $T$.

The 3d method, which we use in this work, uses a three-dimensional (timeslice source) sequential propagator through the interpolating field $\phi^\dagger_H$, as shown in Fig.~\ref{fig:3d_method}.
Using the 3d method, the source for the sequential solve only includes the phase factor $e^{i \vec{p}_H \cdot \vec{y}}$ and the $\gamma_5$ matrix associated with the interpolating field $\phi^\dagger_H$.
For a single value of $t_H$, one calculates the three-point function $C_{\mu \nu}(t_{\rm em}, t_H)$ for all values of $t_{\rm em}$. 
This implies one can calculate both $I^>_{\mu \nu}(t_H, T)$ and $I^<_{\mu \nu}(t_H, T)$ for all $T$. 
Crucially, because $C_{\mu \nu}(t_{\rm em}, t_H)$ is independent of $E_\gamma$, calculating $I_{\mu \nu}(t_H, T)$ for different photon virtualities is done by simply changing the value of $E_\gamma$ in the $e^{E_\gamma t_{\rm em}}$ factor in the integrand of $I_{\mu \nu}(t_H, T)$. 
In other words, using the 3d method, one can calculate all photon virtualities for zero additional propagator solves relative to the real photon decay.
When using the 3d method, this also applies to the spectral reconstruction method \cite{Frezzotti:2023nun}, where the photon virtuality can be changed by changing the kernel.

Furthermore, if one uses point sources, as done in this work, for a fixed value of $\vec{p}_H$, one can also calculate the three-point function for all values of $\vec{p}_\gamma$ for free.
Lastly, as discussed in Ref.~\cite{Giusti:2023pot}, the propagators required to calculate the three-point function with the weak current fixed to the origin in Eq.~\eqref{eq:three_point} can be reused to calculate the three-point function with the EM current fixed to the origin in Eq.~\eqref{eq:C3_EM}.

\begin{table}
    \centering
    \begin{tabular}{ccccccc}
        \hline \hline
        Source & & 3d & & 4d \\
        \hline
        point & & $2(1+N_{t_H}N_{p_H})$ & & $2(1+4N_T N_{p_\gamma} N_{E_\gamma})$ 
        \\ 
        $\mathbb{Z}_2$ wall & & $2(1+N_{t_H} N_{p_H}+N_{p_H}N_{p_\gamma})$ & & $2(1+4 N_T N_{p_\gamma}N_{E_\gamma}+N_{p_\gamma} N_{p_H})$ 
        \\
        \hline\hline
    \end{tabular}
    \caption{Number of propagator solves required to calculate $I_{\mu \nu}(t_H, T)$ using the 3d and 4d methods on a single configuration for a single initial source as a function of the number of meson momenta $N_{p_H}$, number of photon momenta $N_{p_\gamma}$, number of photon energies $N_{E_\gamma}$, number of source-sink separations $N_{t_H}$ for the 3d method and number of integration range values $N_T$ for the 4d method. The number of propagator solves for the 3d method is independent of the number of photon energies $N_{E_\gamma}$, which implies one can calculate all values of photon virtuality for zero additional solves.}
    \label{tab:num_solves}
\end{table}

\section{Lattice parameters}
\label{sec:lattice_setup}

In this work, we perform calculations on a single gauge-field ensemble, namely, one of the ``24I'' ensembles generated by the RBC-UKQCD collaboration~\cite{RBC:2010qam}.
This ensemble has $N_s=24$ sites in the spatial dimensions and $N_T=64$ sites in the Euclidean time dimension, was generated with 2+1 flavors of domain-wall fermions (DWF) and the Iwasaki gauge action, has an inverse lattice spacing $a^{-1} = 1.785(5)$ GeV and pion mass of $m_\pi = 340 (1)$ MeV~\cite{RBC:2014ntl}.
We use the same DWF action for the valence strange quarks as the sea quarks, except that we use the physical bare mass $am_s^{\rm val} = 0.0323$.
For the charm valence quarks, we use a M\"{o}bius DWF action with Stout-smeared gauge links and with a nearly physical charm-quark mass of $am_c = 0.6$ \cite{Boyle:2018knm}.
We use local currents as in Ref.~\cite{Giusti:2023pot}.
We perform calculations on $N_{\rm cfg}=25$ configurations, using either $\mathbb{Z}_2$ random-wall or point sources.
We employ all-mode-averaging, and for each configuration we use 1 exact and 64 sloppy samples.
More details of the calculation can be found in Ref.~\cite{Giusti:2023pot}.

\begin{figure}
    \centering
    \includegraphics[width=0.6\linewidth]{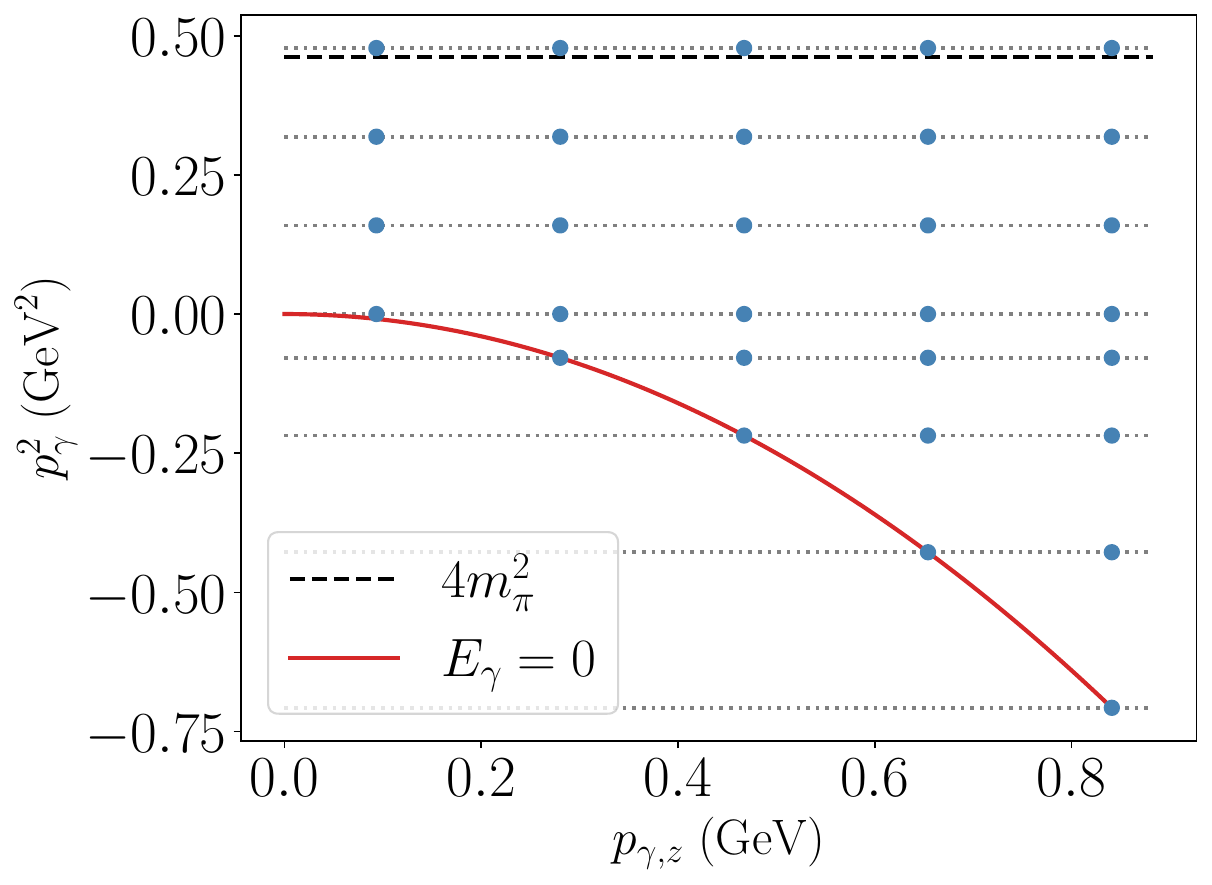}
    \caption{
    Kinematic points studied in this work. 
    The blue points indicate the 30 kinematic points $(p_{\gamma,z}, p_\gamma^2)$ we study.
    The red line corresponds to the theoretically smallest virtuality achievable obtained by setting $E_\gamma=0$. 
    The horizontal dotted gray lines show the 8 lines of constant photon virtuality we study, including the real photon case, and the horizontal black dashed line indicates the maximum virtuality $p_\gamma^2 = 4m_\pi^2$ that can be studied using lattice QCD. 
    }
    \label{fig:kinematics}
\end{figure}

All calculations are performed in the rest frame of the $D_s$ meson with $\vec{p}_{D_s} = (0,0,0)$ and photon momentum aligned with the z-axis $\vec{p}_\gamma = (0,0,p_{\gamma,z})$.
We performed the calculation at 30 different photon 4-momentum, which are visualized in Fig.~\ref{fig:kinematics}.
We perform the calculation at 5 values of $ap_{\gamma,z} = \frac{2\pi}{N_s}(0.2, 0.6, 1.0, 1.4, 1.8)$ and average over positive and negative $p_{\gamma,z}$ (we discuss the method that allows us to study momenta that are not integer multiples of $2\pi/N_s$ in the next section).
In addition to the real photon case $p_\gamma^2=0$, for each $p_{\gamma,z}$ we study both positive and negative virtuality.
For positive virtuality, we use 3 values of positive virtuality equally spaced between 0 the threshold $4m_\pi^2$ (with $m_\pi \approx 340$ MeV).
When considering negative virtuality, the theoretical minimum value that one can achieve is by setting $E_\gamma = 0$.
To set the lines of constant virtuality, we choose the 4 values associated with setting $E_\gamma=0$ for all but the smallest photon momentum, \emph{i.e.}, $-a^2p_\gamma^2 = (\frac{2\pi}{N_s})^2\{(0.6)^2, (1.0)^2, (1.4)^2, (1.8)^2\}$.
For each $p_{\gamma, z}$, we perform the calculation at all allowed values of negative virtuality in this list; the kinematic points we study are shown in Fig.~\ref{fig:kinematics}.
In physical units, the largest negative virtuality we study is $p_\gamma^2 \sim -0.7$ GeV$^2$.
Lastly, for each kinematic point, we perform the calculation for two values of source-sink separation $-t_H/a \in \{9, 12\}$.

\section{Improved estimators}
\label{sec:imp_estimators}

In this section, we briefly review the improved estimators for $C_{\mu \nu}^{V}(t_{\rm em}, t_H)$ and $C_{\mu \nu}^{V, {\rm EM}}(t_{\rm em}, t_H)$ studied in Ref.~\cite{Giusti:2023pot}.
Here we only discuss the high-level details and the qualitative improvements of each method.
For a detailed discussion, we refer the reader to the original reference.

The first improvement is the so-called infinite-volume approximation.
Using this approximation, we calculate correlation functions projected to arbitrary momenta (not restricted to integer multiples of $2\pi/L$) with errors exponentially suppressed in the finite volume.
At a high level, this approximation relies on the fact that correlation functions decrease exponentially with the spatial separation of the operators.
This exponential decay can then be used to bound the difference between the infinite-volume and finite-volume correlation functions to also be exponentially suppressed in the volume.
In this way, we can study arbitrarily small values of $x_\gamma$ in the rest frame of the initial-state meson $H$.
Without this approximation, achieving small values of $x_\gamma$ would require working in the moving frame of the initial state meson $H$, which results in an increase in statistical noise~\cite{Giusti:2023pot}.

One caveat of the infinite-volume approximation is that it requires control of the distance of the two currents.  It is therefore most naturally implemented using point sources,
which generally have worse statistical performance compared to using $\mathbb{Z}_2$-wall sources.  It was shown that this limitation can be overcome by using ratios of three-point functions calculated with both point and noise sources.
If we denote by $C^{\rm point}_{3, \mu \nu}(\vec{p}_\gamma, t, t_H)$ the three-point function (where $t$ could be $t_{\rm em}$ or $t_W$) calculated with photon-momentum $\vec{p}_\gamma$ in the rest frame of $H$, the improved estimator we use is  
\begin{align}
    C^{\rm improved}_{3, \mu \nu}(\vec{p}_\gamma, t, t_H) = C^{\rm point}_{3, \mu \nu}(\vec{p}_\gamma, t, t_H) \frac{C^{\mathbb{Z}_2}_{3, \mu \nu}(\vec{p}^*, t, t_H)}{C^{\rm point}_{3, \mu \nu}(\vec{p}^*, t, t_H)},
    \label{eq:ratio_C3}
\end{align}
where $\vec{p}^*$ is some momentum allowed by periodic boundary conditions, and $C^{\mathbb{Z}_2}_{3, \mu \nu}(\vec{p}^*, t, t_H)$ is the three-point function calculated using $\mathbb{Z}_2$-wall sources.
Note that one must calculate the expectation values of the individual correlation functions before taking the ratio.
Because $C_{3,\mu\nu}^{V}$ and $C_{3,\mu\nu}^{V, \rm{EM}}$ have zero expectation value when $\vec{p}_\gamma=0$, we choose $a\vec{p}^*=\frac{2\pi}{N_s}(0,0,1)$.
For each value of $t$, we compare the statistical uncertainty of the improved and original three-point function and use the result with the smaller uncertainty.

Another improvement we employ is to average over $\pm \vec{p}_\gamma$. 
While this may at first appear to be a trivial improvement, it was shown that doing so results in an exact cancellation of a pure noise contribution to the weak-vector component of the three-point function.
Averaging therefore results in dramatic noise reductions, in particular at small photon momentum~\cite{Giusti:2023pot}.

\section{Fit methods to control unwanted exponentials}
\label{sec:fit_methods}

In this section, we describe the procedure used to extrapolate to infinite integration range $T$ and source-sink separation $t_H$.
The methods we use are similar to those in Ref.~\cite{Giusti:2023pot}, with slight modifications.
All expressions in this section are written in the rest frame of the initial state meson $H$ with $\vec{p}_H = (0,0,0)$.

We start by discussing the behavior of the unwanted exponentials as one varies $t_H$ and $T$ for the different time orderings.
Consider first the unwanted exponentials associated with the $t_{\rm em}<0$ and $t_W>0$ time orderings.
Assuming we have achieved ground-state saturation by taking $t_H$ large and negative, the unwanted exponentials are of the form
\begin{align}
    t_{\rm em}<0&: \quad e^{-(E_\gamma+E_{n,\vec{p}_\gamma}-m_H)T}, \quad {\rm where} \quad 0 < T < -t_H,
    \\
    t_W>0&: \quad e^{-(E_\gamma+E_{n,\vec{p}_\gamma}-m_H)T}, \quad {\rm where} \quad 0 < T < aN_T/2,    
\end{align}
where $n$ is the index of the intermediate states in the spectral decompositions, and we have explicitly written the ranges of $T$ one has access to, where $aN_T$ is the temporal extent of the lattice.
These expressions imply that the unwanted exponentials decay more quickly for larger values of $E_\gamma$ and $\vec{p}_\gamma$, and more slowly for smaller $E_\gamma$ and $\vec{p}_\gamma$.

For the $t_{\rm em}>0$ and $t_W<0$ time orderings, the associated unwanted exponentials are (assuming ground-state saturation)
\begin{align}
    t_{\rm em}>0&: \quad e^{(E_\gamma-E_{n,\vec{p}_\gamma})T}\,, \quad {\rm where} \quad 0 < T < aN_T/2,
    \\
    t_W<0&: \quad e^{(E_\gamma-E_{n,\vec{p}_\gamma})T}, \quad {\rm where} \quad 0 < T < -t_H\,.
\end{align}
In general, we see that these terms decay more slowly with $T$ for large $E_\gamma$ and large $\vec{p}_\gamma$, and actually start to grow with $T$ when $p_\gamma^2 > 4m_\pi^2$.

For both time orderings, we see that there are kinematic regions where controlling unwanted exponentials will be difficult, requiring either taking long times $T$ or using more complicated fit forms.
This problem is particularly difficult for the $t_{\rm em}<0$ and $t_W<0$ datasets, as one cannot integrate past the interpolating field, \emph{i.e.}, $T < -t_H$.
Furthermore, because the excited-state contamination increases as one integrates towards $\phi^\dagger_H$, one must simultaneously take $T \gg 0$ and $T+t_H \gg 0$.
While it was found that such extrapolations were possible for real photon decays~\cite{Giusti:2023pot}, certain kinematic regimes are more difficult to deal with when considering off-shell photons.
To simplify the analysis and limit the possibility of uncontrolled systematic uncertainties, 
we choose to only include the $t_{\rm em}>0$ and $t_W>0$ data in our analysis.
We leave including the $t_{\rm em}<0$ and $t_W<0$ datasets in the analysis for future work.

Following Ref.~\cite{Giusti:2023pot}, rather than first fitting the time orderings of $I_{\mu \nu}$ and $I_{\mu \nu}^{\rm EM}$ and then taking linear combinations of the fit results to calculate $F_V$, we choose to fit the form factor as a function of $T$ and $t_H$ directly.
Furthermore, we choose to fit data associated with the charm $(c)$ and strange $(s)$ quark components of the EM current separately.
This is done for two reasons.
First, this is actually necessary for the $t_{\rm em}>0$ and $t_W<0$ datasets as different intermediate states contribute to the $(c)$ and $(s)$ components.
Second, the individual charm $F_V^{(c)}$ and strange $F_V^{(s)}$ components of the form factor are themselves phenomenologically relevant, see, \emph{e.g.}, Refs~\cite{Donald:2013sra, Pullin:2021ebn}. 
Note that we define the quark components to include the EM quark charge factors such that $F_V = F_V^{(c)} + F_V^{(s)}$.
As an example of the notation, in the rest frame of the initial state meson with photon momentum $\vec{p}_\gamma=(0,0,p_{\gamma,z})$, we fit
\begin{equation}
    F_V^{(q),>}(t_H, T) = \frac{1}{2 p_{\gamma,z}} (I^{(q),>}_{21}(t_H, T)-I^{(q),>}_{12}(t_H, T))
\end{equation}
where the superscript $q = c,s$; we use similar notation for the other time ordering and correlation function.

The integrals over $t_{\rm em}$ and $t_W$ are performed using the trapezoid rule, with the $t_{\rm em}=0$ $(t_W=0)$ contribution split evenly between the two time orderings.
Furthermore, because $F_V(t_H, T+a)$ is the sum of $F_V(t_H, T)$ and an exponentially decaying contribution, the data at different $T$ are highly correlated, leading to instabilities in fully correlated fits.
To avoid these instabilities, we follow Ref.~\cite{Giusti:2023pot} and instead perform uncorrelated fits and estimate the statistical uncertainties using a jackknife procedure.
For each set of fits, we perform fits with many different minimum values of $T$ and pick the value where the form factor extracted from the fit is stable.
To help stabilize the fits, we place a Gaussian prior in the excited-state energy gap $\Delta E$ between the ground state and the first excited state created by the interpolating field $\phi^\dagger_H$ by fitting the associated two-point correlation function to a two-exponential fit form; the prior value is chosen as the central result of the fit and the prior width is chosen as the statistical uncertainty of the fit result scaled by a factor of $1.5$.

We now consider the quantum numbers of the intermediate states that contribute to the spectral decomposition of the different time orderings\footnote{Note that we consider the quantum numbers assuming continuum and infinite-volume QCD.}.
For the $t_{\rm em}>0$ data, the intermediate states are flavorless, and the lowest-energy state is in principle a two-pion state. 
However, as we did not include disconnected diagrams in our calculation, this state does not actually contribute; because we study the $D_s$, the intermediate states contributing to the correlation function involving the $s$ and $c$ quark components of the EM current can only be composed of flavorless combinations of $(s,\bar{s})$ and $(c,\bar{c})$ quarks, respectively.
This implies that the intermediate states contributing to $F_V^{(s),>}$ and $F_V^{(c),>}$ are different, and so, as previously discussed, we fit them separately.
This also implies that the largest virtuality we can study is larger than $4m_\pi^2$, although the results should not be taken as physical because we currently neglect the disconnected diagrams.

The most general fit form we consider~\cite{Giusti:2023pot} includes a term to account for the unwanted exponential associated with the lowest-energy intermediate state and the lowest-energy excited state created by the interpolating field,
\begin{align}
    F_V^{(q),>}(t_H, T) &= F_V^{(q),>} + B_{(q)}^{>} e^{(E_\gamma - E_{(q)}^>)T} + C^{>}_{q} e^{\Delta E t_H}\,.
\end{align}

\begin{figure}
    \centering
    \includegraphics[width=0.9\linewidth]{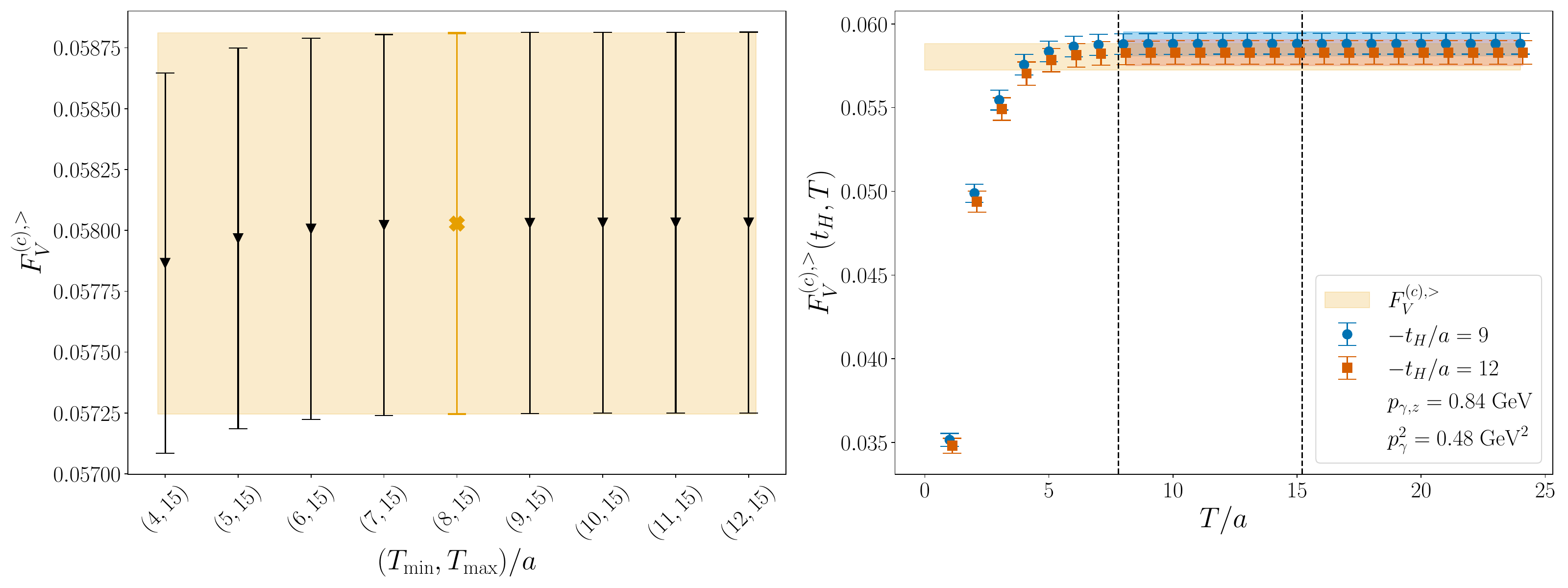}
    \caption{Left: Value of $F_V^{(c),>}$ determined from fits to $F_V^{(c),>}(t_H,T)$ as a function of the minimum value of integration range $T_{\rm min}$ of data points included in the fit.
    As $T_{\rm min}$ is increased, the fit result approaches a constant, demonstrating the unwanted exponentials are well controlled.
    The yellow data point and error band show the chosen fit range.
    Right: Fit result of the chosen fit range on top of the data. The vertical black dashed lines indicate the fit range used. The blue band is the fit function with $-t_H/a=9$, the orange band is the fit function with $-t_H/a=12$, and the yellow error band with $-t_H/a=\infty$; the yellow band on the left figure is the same as the yellow band on the right figure.
    The results shown are for photon momentum  $p_{\gamma,z}=0.84$ GeV ($ap_{\gamma,z} = (1.8)\frac{2\pi}{N_s}$) and photon virtuality $p_\gamma^2 = 0.48$ GeV, which corresponds to $p_\gamma^2 \sim 4m_\pi^2$.}
    \label{fig:FV_c_weak_pos_vs_T}
\end{figure}

Because intermediate states for the $q=c$ component are composed of at least one charm and one anti-charm quark, the energy $E_{(c)}^>$ is large and the unwanted exponentials decay rapidly with $T$.
Looking at Fig.~\ref{fig:FV_c_weak_pos_vs_T} we see that, even in the worst-case kinematic point we study, with $ap_{\gamma,z} = (1.8)\frac{2\pi}{N_s}$ and $p_\gamma^2 \sim 4m_\pi^2$, the unwanted exponentials decay rapidly and one can simply fit to a constant in $T$.
For this reason, we set $B_{(c)}^>=0$ for all kinematic points considered in this work.

\begin{figure}
    \centering
    \includegraphics[width=0.8\linewidth]{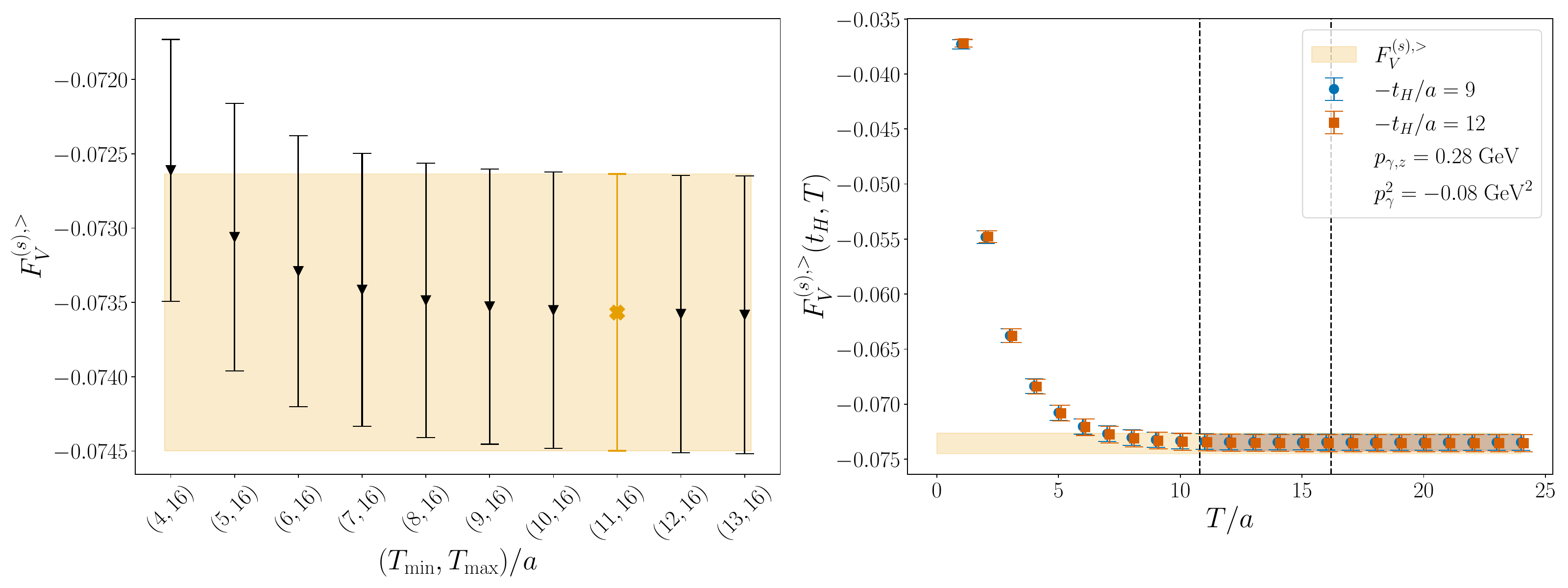}
    \includegraphics[width=0.8\linewidth]{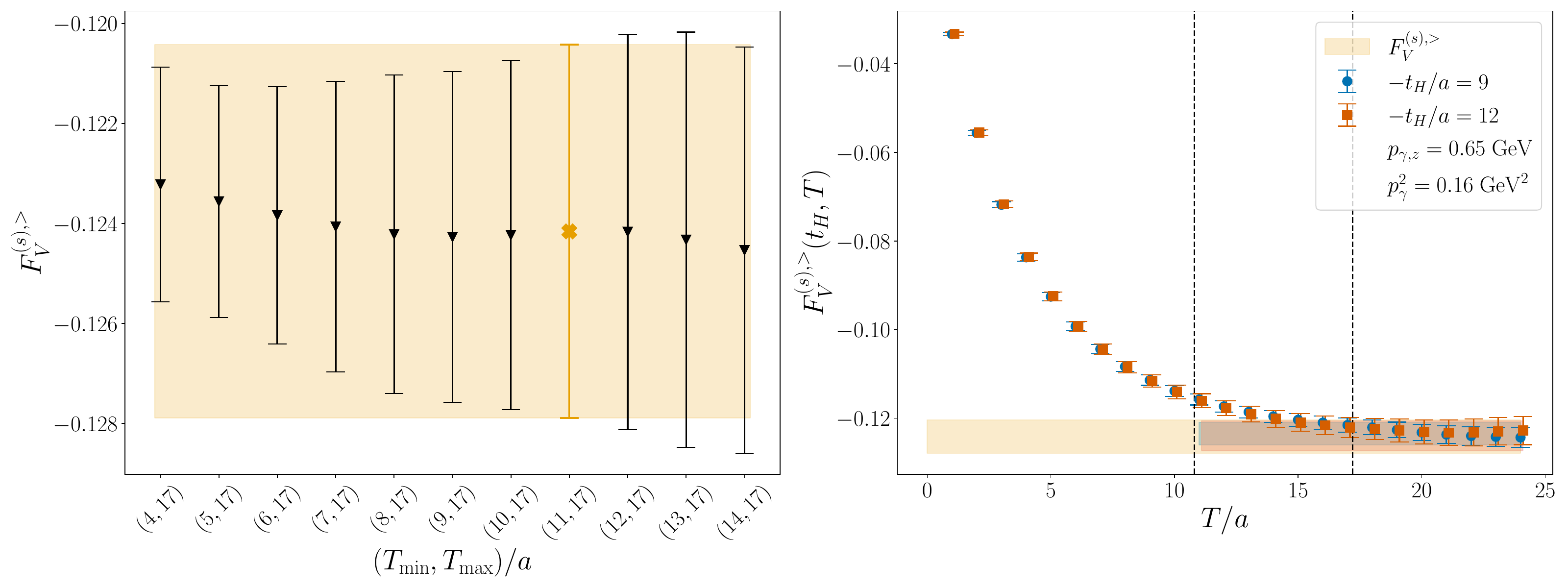}
    \includegraphics[width=0.8\linewidth]{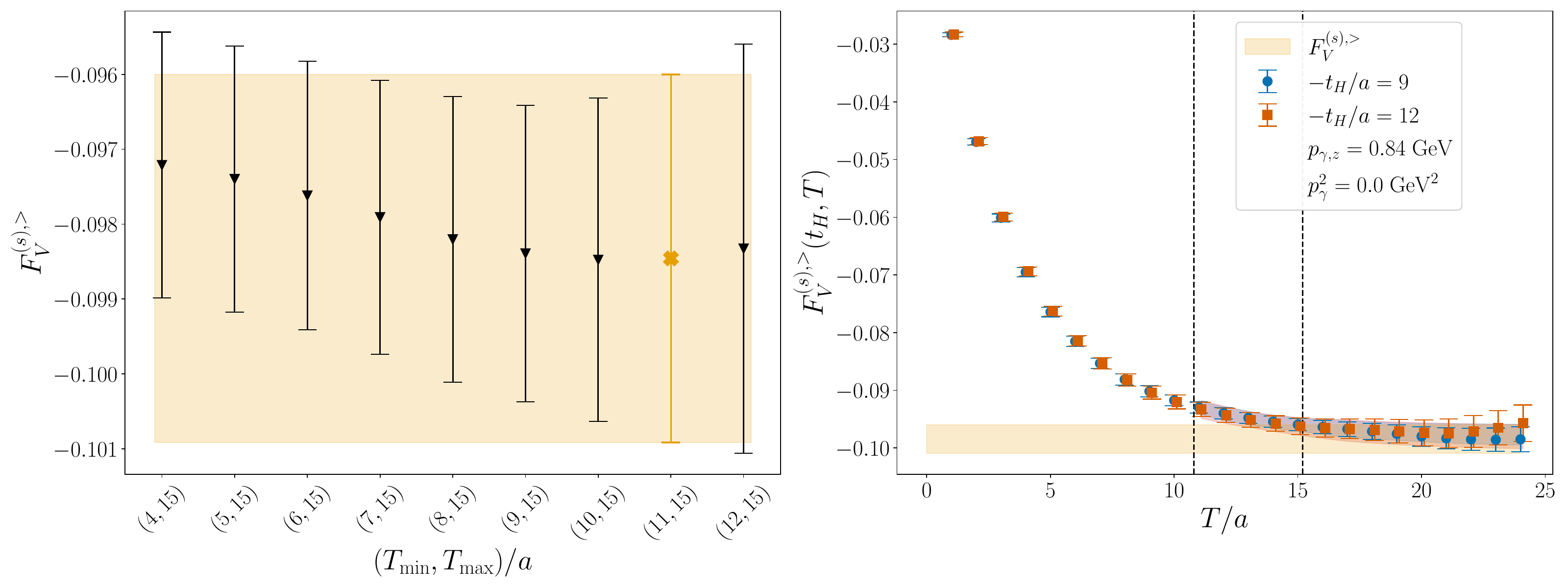}
    \includegraphics[width=0.8\linewidth]{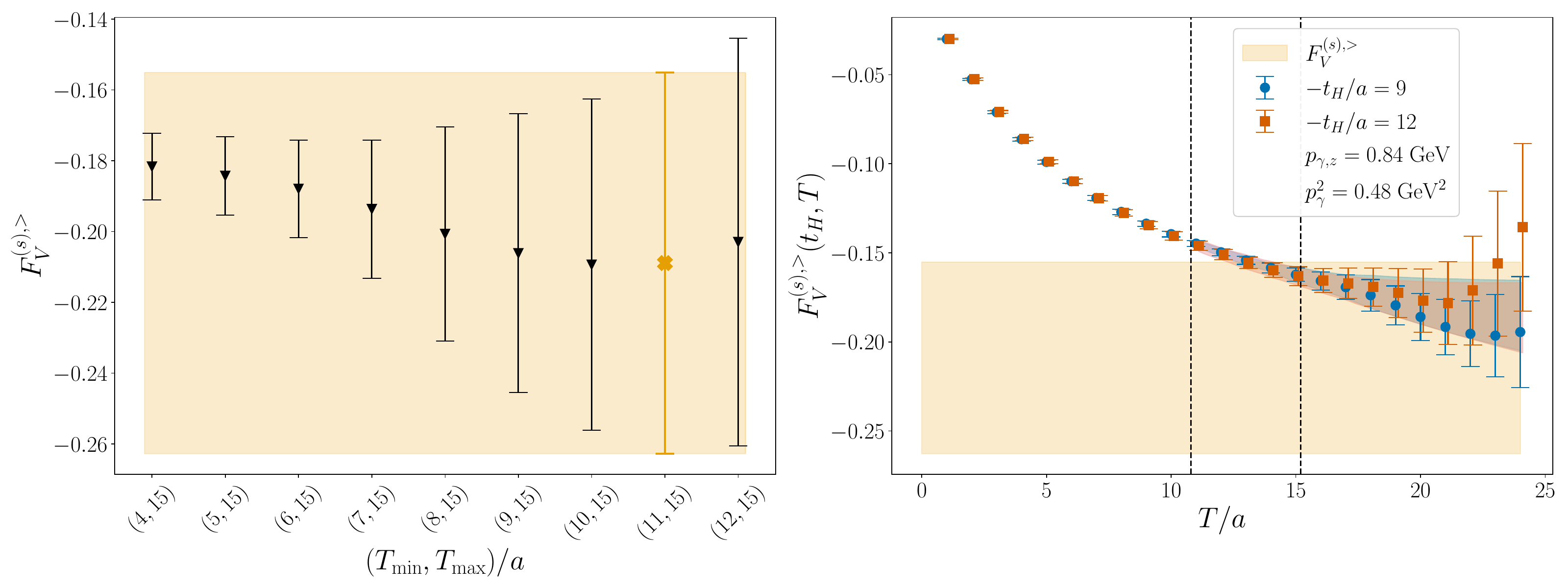}
    \caption{Same as Fig.~\ref{fig:FV_c_weak_pos_vs_T} but for $F_V^{(s),>}$.
    The left figure shows the results of the form factor as a function of the minimum fit range, and the associated fit result on top of the data are shown in the same row in the right column. 
    Different rows correspond to different kinematic points.}
    \label{fig:fits_FV_s_weak_pos}
\end{figure}

We now discuss fits to the $F_V^{(s),>}(t_H, T)$ data.
Because the intermediate states are composed of strange quarks, they have lower energies and the unwanted exponentials decay more slowly than in the $F_V^{(c),>}(t_H, T)$ case.
In Fig.~\ref{fig:fits_FV_s_weak_pos}, we show example fit results for $F_V^{(s),>}$ at different kinematic points.
At low photon energy and momentum, the unwanted exponentials decay quickly with $T$ and we perform fits with the $B_{(s)}^>=0$.
As the photon energy and momentum are increased, the unwanted exponentials decay more slowly and we include the term proportional to $B_{(s)}^>$ in the fits.
One important property of these correlation functions is that, for large enough $p_{\gamma,z}$ and $p_\gamma^2$, there is a threshold $T$ for which the signal-to-noise ratio decays exponentially with increasing $T$~\cite{Frezzotti:2023ygt}.
If not properly taken into account, this behavior can lead to underestimating statistical uncertainties.
An example of this behavior is shown in the bottom row of Fig.~\ref{fig:fits_FV_s_weak_pos} for $ap_{\gamma,z}=(1.8)\frac{2\pi}{N_s}$ and $p_\gamma^2 \sim 4m_\pi^2$. 
Looking at the associated stability plot, one might conclude that $T_{\rm min}/a$ can be chosen as small as $4-6$.
Doing so, however, would result in an underestimation of the statistical uncertainties.
To see this, first recall that, at fixed photon momentum $p_{\gamma,z}$, the unwanted exponentials for this time ordering decay more slowly with increasing virtuality $p_\gamma^2$.
This implies that one should choose $T_{\rm min}$ at least as large as that chosen for the real photon decay at the same photon momentum.
As shown in Fig.~\ref{fig:fits_FV_s_weak_pos}, a value of $T_{\rm min}/a=11$ was chosen for $ap_{\gamma,z}=(1.8)\frac{2\pi}{N_s}$ and $p_\gamma^2 = 0$, and so we use the same value for the non-zero virtuality case. 
We use this strategy for all fits to $t_{\rm em}>0$ data.

\begin{figure}
    \centering
    \includegraphics[width=0.98\linewidth]{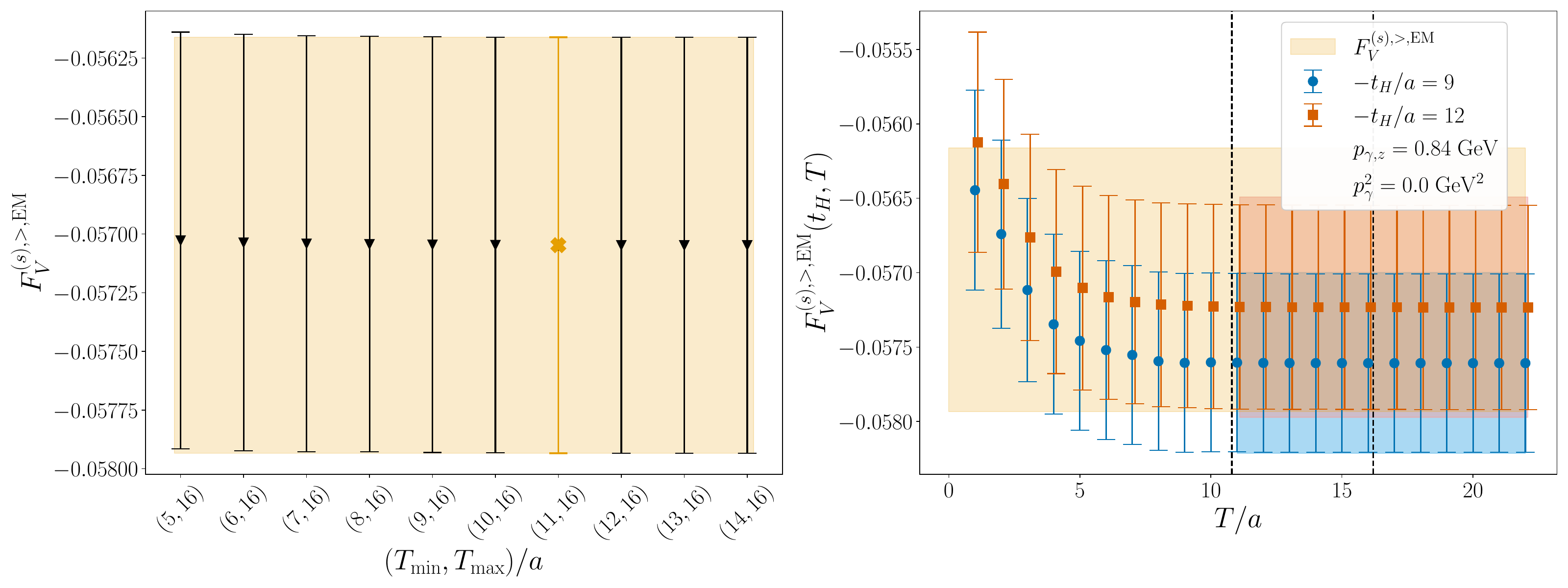}
    \includegraphics[width=0.98\linewidth]{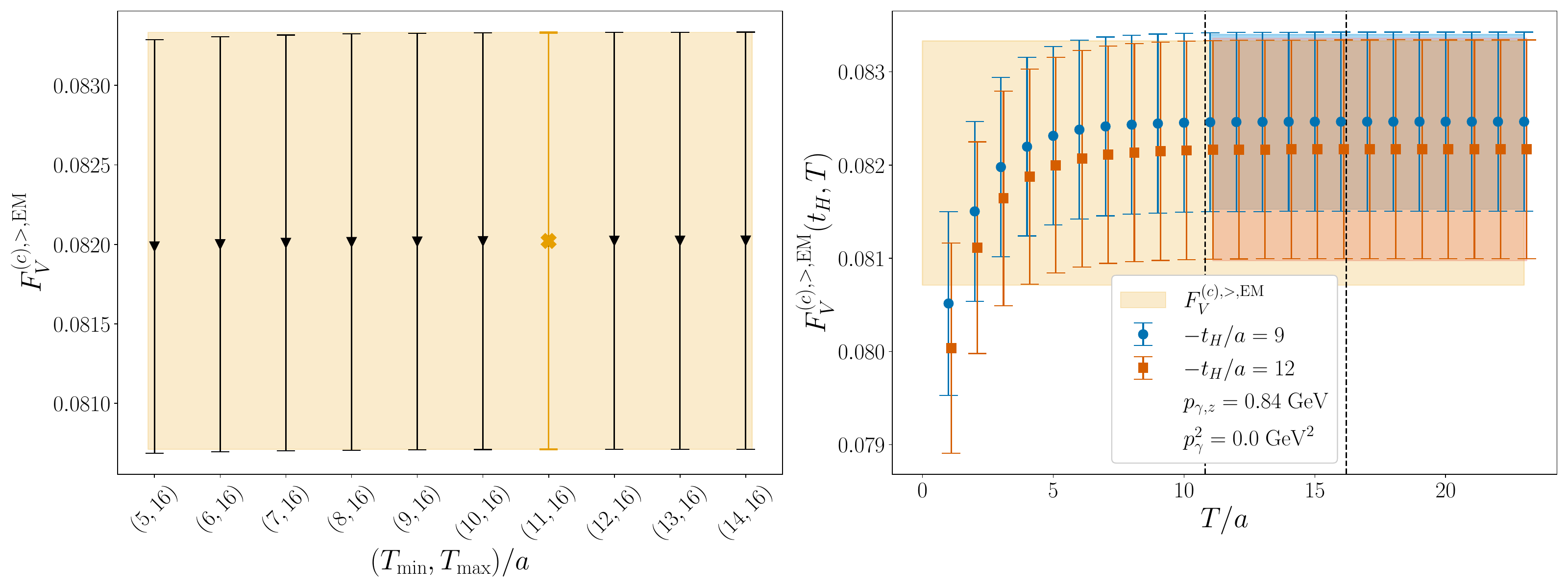}
    \caption{Same as Fig.~\ref{fig:FV_c_weak_pos_vs_T}, but the top and bottom rows show fits to $F_V^{(s),>,{\rm EM}}(t_H, T)$ and $F_V^{(c),>,{\rm EM}}(t_H, T)$, respectively.
    The kinematic point shown is the real photon case with $p_{\gamma,z} = (1.8)\frac{2\pi}{L}$, which is where the unwanted exponentials are expected to decay most quickly for this time ordering.}
    \label{fig:fits_FV_s_and_c_em_pos_best_case}
\end{figure}

\begin{figure}
    \centering
    \includegraphics[width=0.98\linewidth]{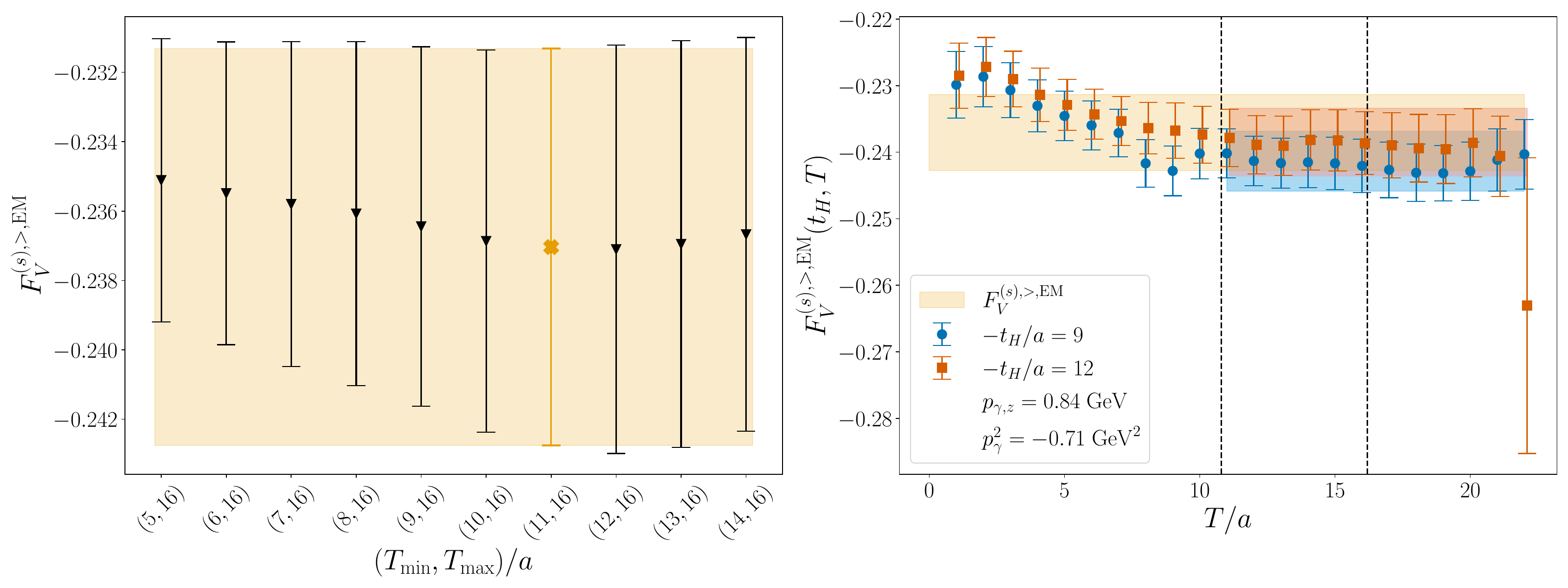}
    \includegraphics[width=0.98\linewidth]{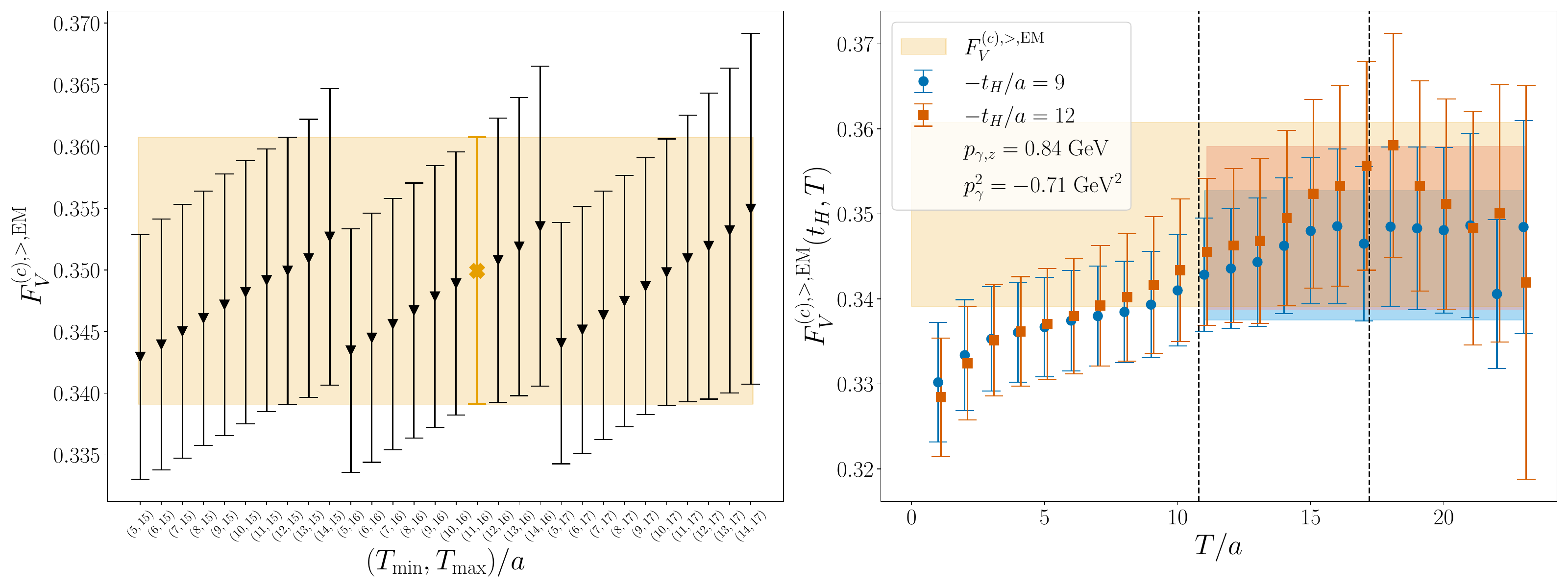}
    \caption{Same as Fig.~\ref{fig:FV_c_weak_pos_vs_T}, but the top and bottom rows show fits to $F_V^{(s),>,{\rm EM}}(t_H, T)$ and $F_V^{(c),>,{\rm EM}}(t_H, T)$, respectively.
    The kinematic point shown is the virtual photon decay with $p_{\gamma,z} = 0.84$ GeV and $p_\gamma^2 = -0.71$ GeV (corresponding to $E_\gamma=0$). 
    While the $F_V^{(s),>,{\rm EM}}(t_H, T)$ data exhibits a clear plateau, the $F_V^{(c),>,{\rm EM}}(t_H, T)$ is more noisy and it is less clear which fit range to choose.
    As explained in the main text, the intermediate states that contribute to the charm and strange quark components are the same, and we therefore choose the fit range for $F_V^{(c),>,{\rm EM}}(t_H, T)$ to be the same as the fit range chosen for the less noisy $F_V^{(s),>,{\rm EM}}(t_H, T)$ data.}
    \label{figig:fits_FV_s_and_c_em_pos_worst_case}
\end{figure}

Turning to the $t_W>0$ data, the intermediate states have the same flavor quantum numbers as the initial state pseudocalar meson $H$.
This is true for both the charm and strange EM current components.
For decays mediated by the weak vector current, the lowest-energy state that contributes is the $H^*$-like state, \emph{i.e.}, for a $D_s \to \gamma^* \ell \nu$ decay this would be the $D_s^*$.
Because we can easily the measure the $D_s^*$ energy $E_{D_s, \vec{p}_\gamma}$ on the lattice, \emph{e.g.} by fitting two-point correlation functions, we can use the method first developed in Ref.~\cite{Tuo:2021ewr} to exactly subtract the unwanted exponential associated with this state.
After subtracting the unwanted exponential associated with the $D_s^*$ state, we observed that, for all kinematic points studied in this work, the data plateaus quickly in $T$, and we use the simple fit form
\begin{align}
    F_V^{(q), >, \text{EM}}(t_H, T) &= F_V^{(q),<} + C^{<,\text{EM}}_{(q)} e^{\Delta E t_H}\,.
\end{align}
Figure~\ref{fig:fits_FV_s_and_c_em_pos_best_case} shows fit results for the best-case scenario when the unwanted exponentials decay most quickly, namely for the real photon case with $ap_{\gamma,z} = (1.8)\frac{2\pi}{N_s}$.
For both $F_V^{(s),>,{\rm EM}}(t_H, T) $ and $F_V^{(c),>,{\rm EM}}(t_H, T)$, the data plateaus quickly with $T$ and the fit results are entirely consistent as one varies $T_{\rm min}$.
As the photon energy is decreased, the unwanted exponentials decay more slowly.
Figure~\ref{figig:fits_FV_s_and_c_em_pos_worst_case} shows fits to $ap_{\gamma,z} = (1.8)\frac{2\pi}{N_s}$ with now $aE_\gamma = 0$.
We see that, while the strange-quark data plateaus in a clear way, the charm-quark data is noisy and it is not obvious which fit ranges to choose.
As a rule of thumb for deciding which fit range to choose for fits to $F_V^{(c),>,{\rm EM}}(t_H, T)$, we use the same fit range chosen for fits to $F_V^{(s),>,{\rm EM}}(t_H, T)$.
This strategy is valid because the same intermediate states contribute to both the strange and charm quark correlation functions, and therefore the unwanted exponentials should become negligible for similar values of $T$.
As a cross-check, we also performed fits to the sum $F_V^{>,{\rm EM}}(t_H, T) = F_V^{(c),>,{\rm EM}}(t_H, T) + F_V^{(s),>,{\rm EM}}(t_H, T)$ directly and found these results to be entirely consistent with first fitting $F_V^{(s),>,{\rm EM}}(t_H, T) $ and $F_V^{(c),>,{\rm EM}}(t_H, T)$ and then adding the fit results.

\section{Vector form factor results at positive and negative photon virtuality}
\label{sec:final_res}
In this section, we show our results for the vector form factor as a function of the photon momentum $p_{\gamma,z}$, energy $E_\gamma$, and virtuality $p_\gamma^2$.
Because we show results for a single gauge field ensemble (the ``24I'' ensemble described in Sec~\ref{sec:lattice_setup}), we focus on the qualitative features of the results.
Additionally, it is important to note that while the disconnected diagrams (which we currently do not include) are expected to be small for the real photon decay (and were recently calculated to be negligible for the $K \to \gamma \ell \nu$ decay with real photons~\cite{DiPalma:2025iud}), it is possible their contribution becomes enhanced as the virtuality approaches $p_\gamma^2 \sim 4m_\pi^2$.

We start by showing results with 2d plots where the x-axis corresponds to $x_\gamma$ and different colored points indicate data at constant virtuality $p_\gamma^2$.
To offer a more rich perspective to view the behavior of the form factor, we also provide 3d plots.
For each data set, we show the same form factor with two different axis; the first shows the dependence on $(p_{\gamma,z}, E_\gamma)$ and the second on $(E_\gamma, p_\gamma^2)$.

\begin{figure}
    \centering
    \includegraphics[width=0.48\linewidth]{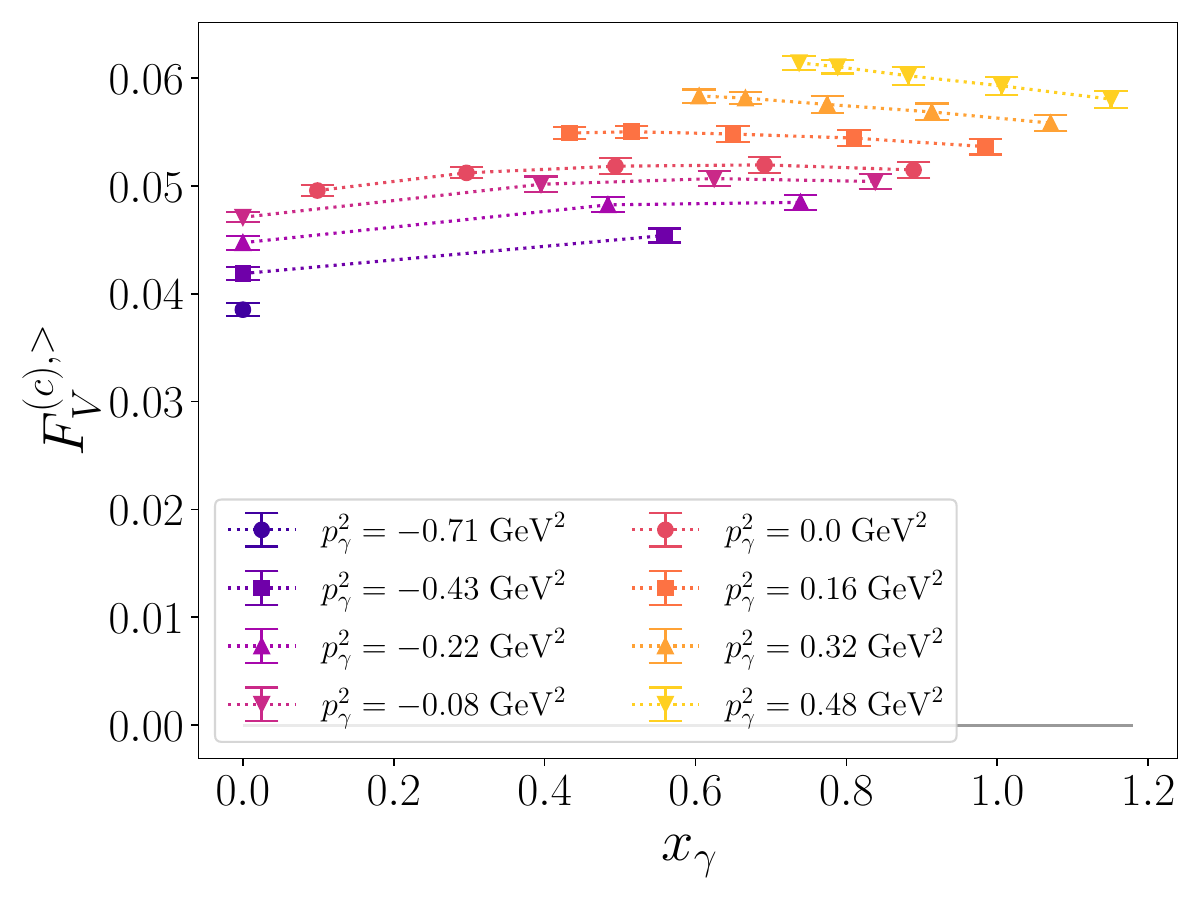}
    \includegraphics[width=0.48\linewidth]{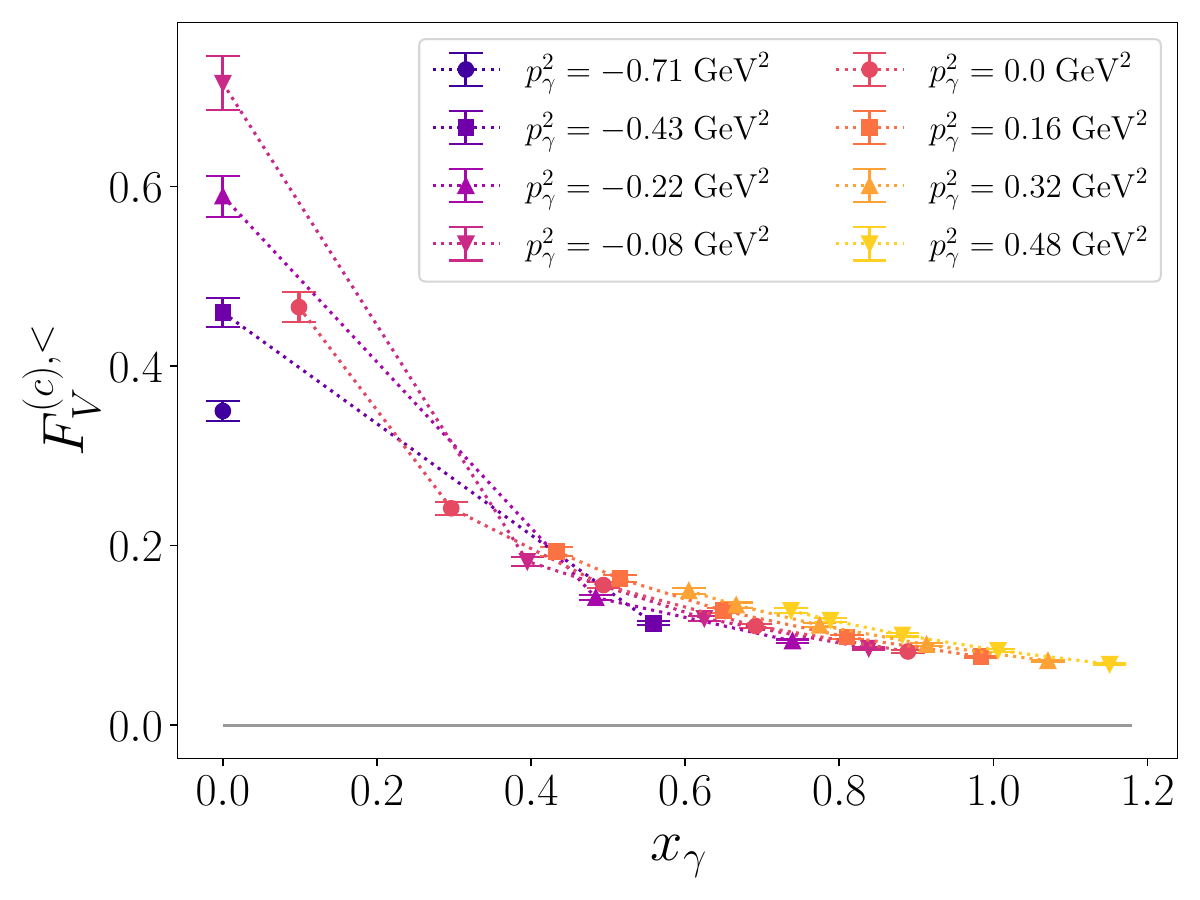}
    \includegraphics[width=0.48\linewidth]{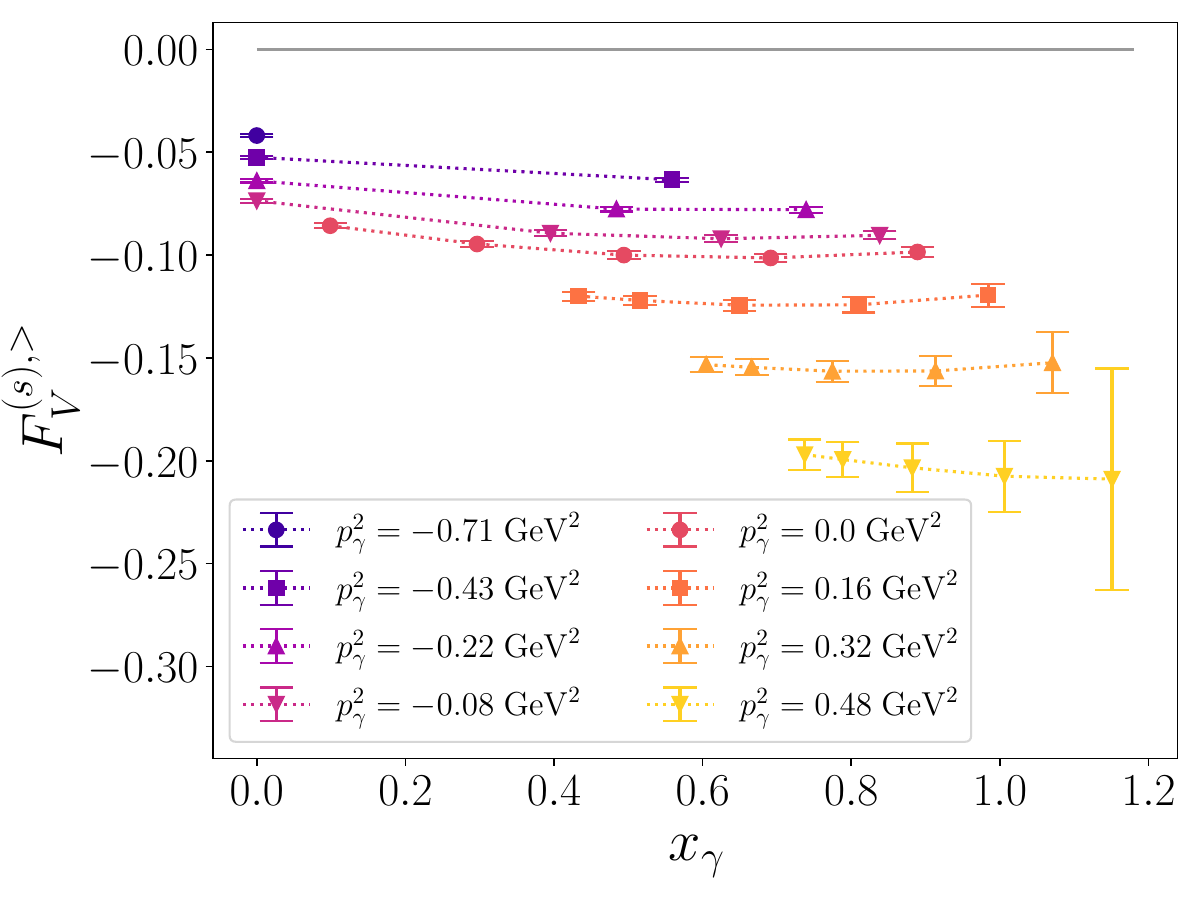}
    \includegraphics[width=0.48\linewidth]{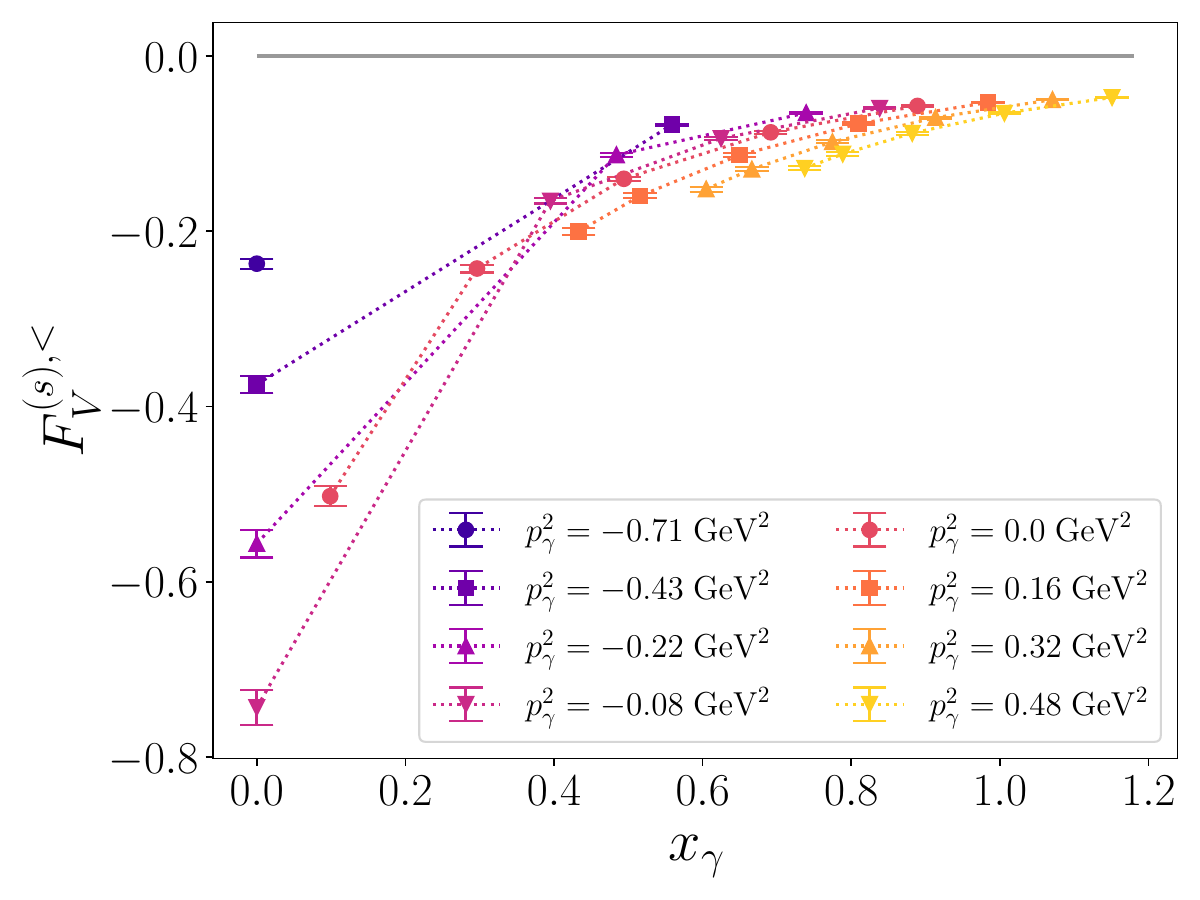}
    \includegraphics[width=0.48\linewidth]{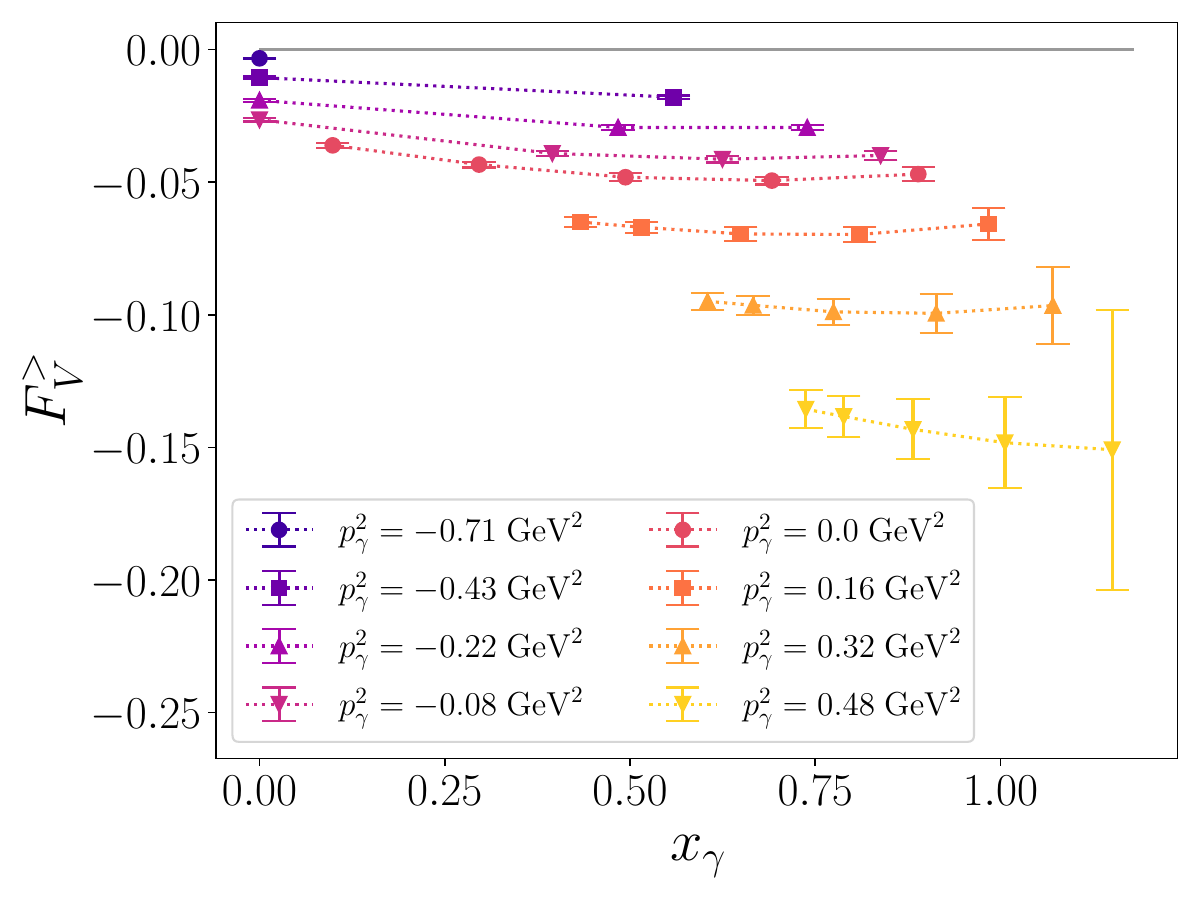}
    \includegraphics[width=0.48\linewidth]{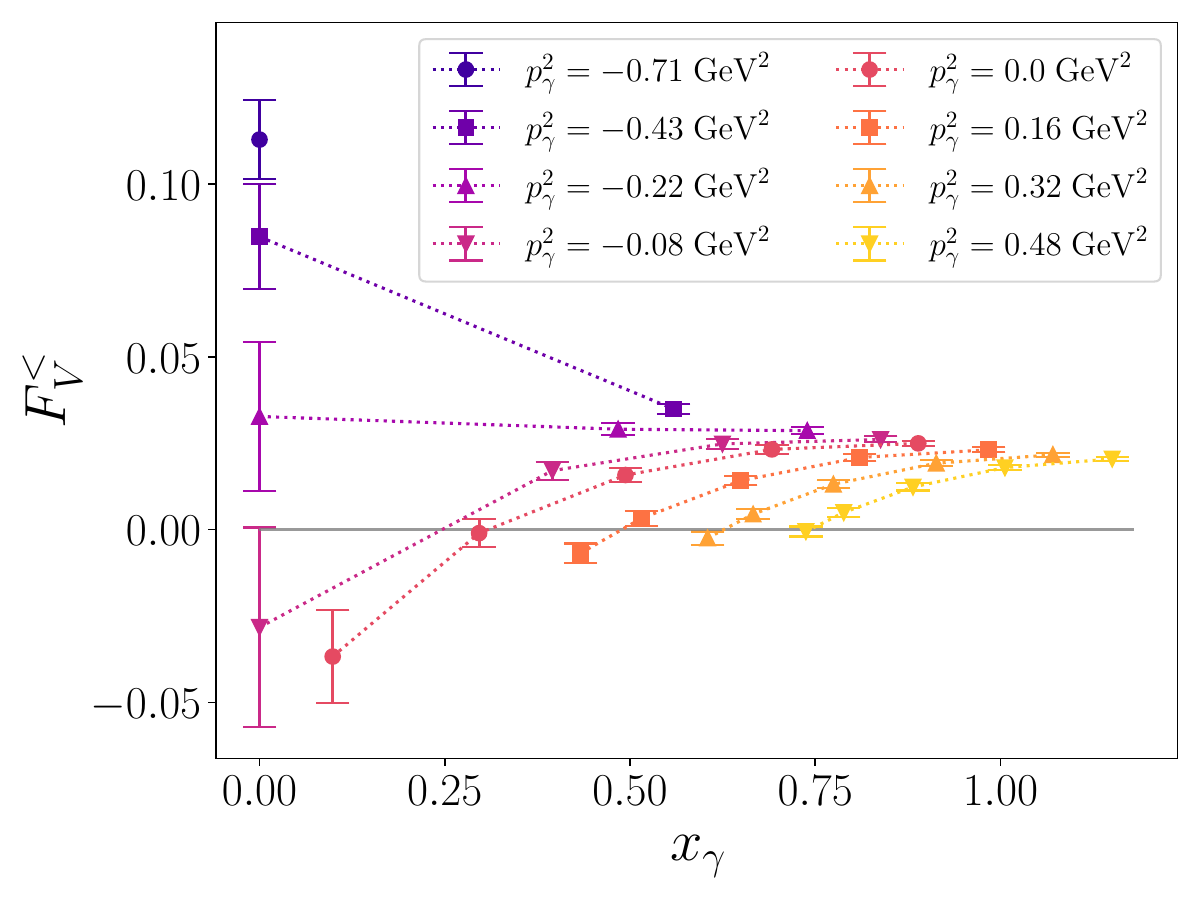}
    \caption{Different components of the vector form factor as a function of $x_\gamma$. Data with different colors and markers correspond to different values of fixed photon virtuality $p_\gamma^2$. The left and right columns show the $F_V^>$ and $F_V^<$ time orderings, respectively. The upper, center, and lower rows show the charm quark component, strange quark component, and the sum of the two, respectively. We observe strong cancellations between the individual charm and strange quark components for both time orderings. Note that the dashed lines connecting different points are meant only to guide the eye.}
    \label{fig:indiv_time_orderings_FV}
\end{figure}

\begin{figure}
    \centering
    \includegraphics[width=0.48\linewidth]{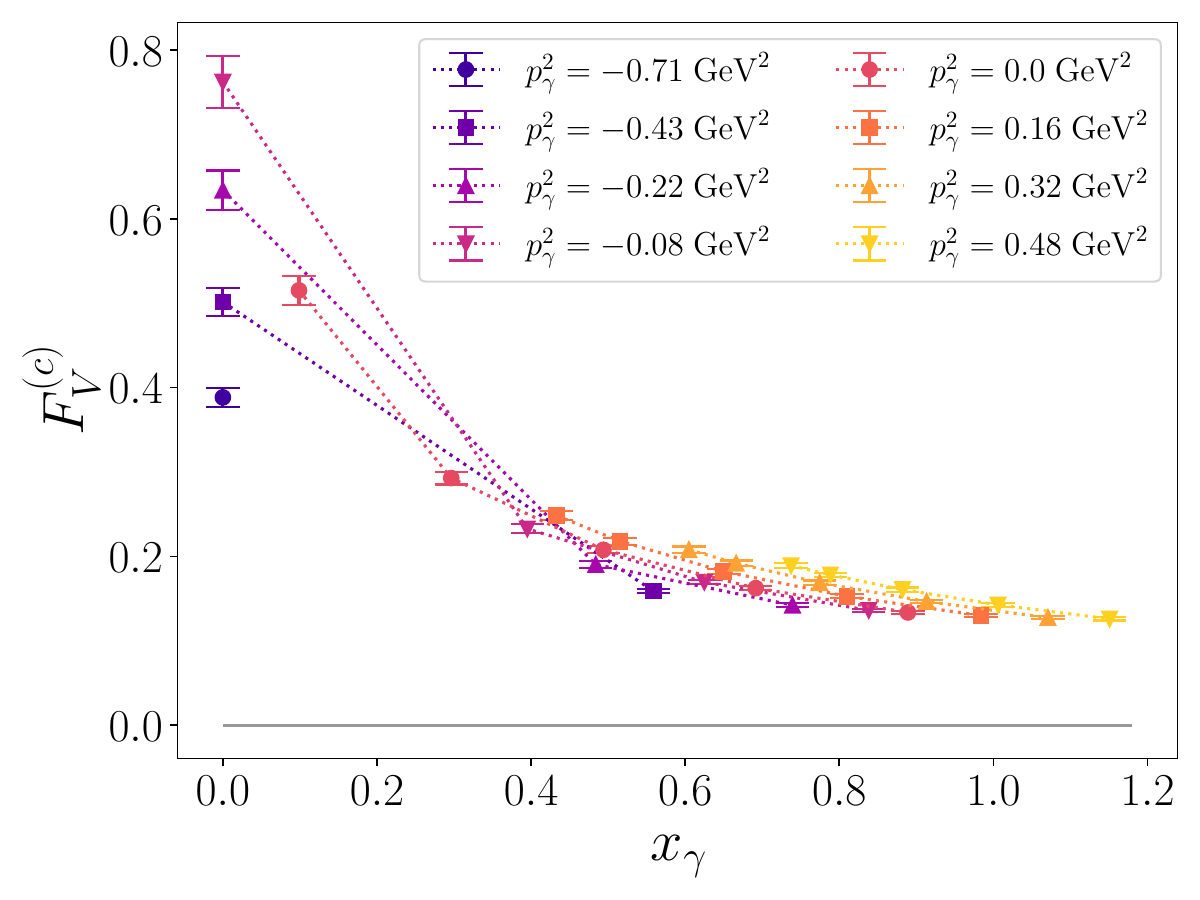}
    \includegraphics[width=0.48\linewidth]{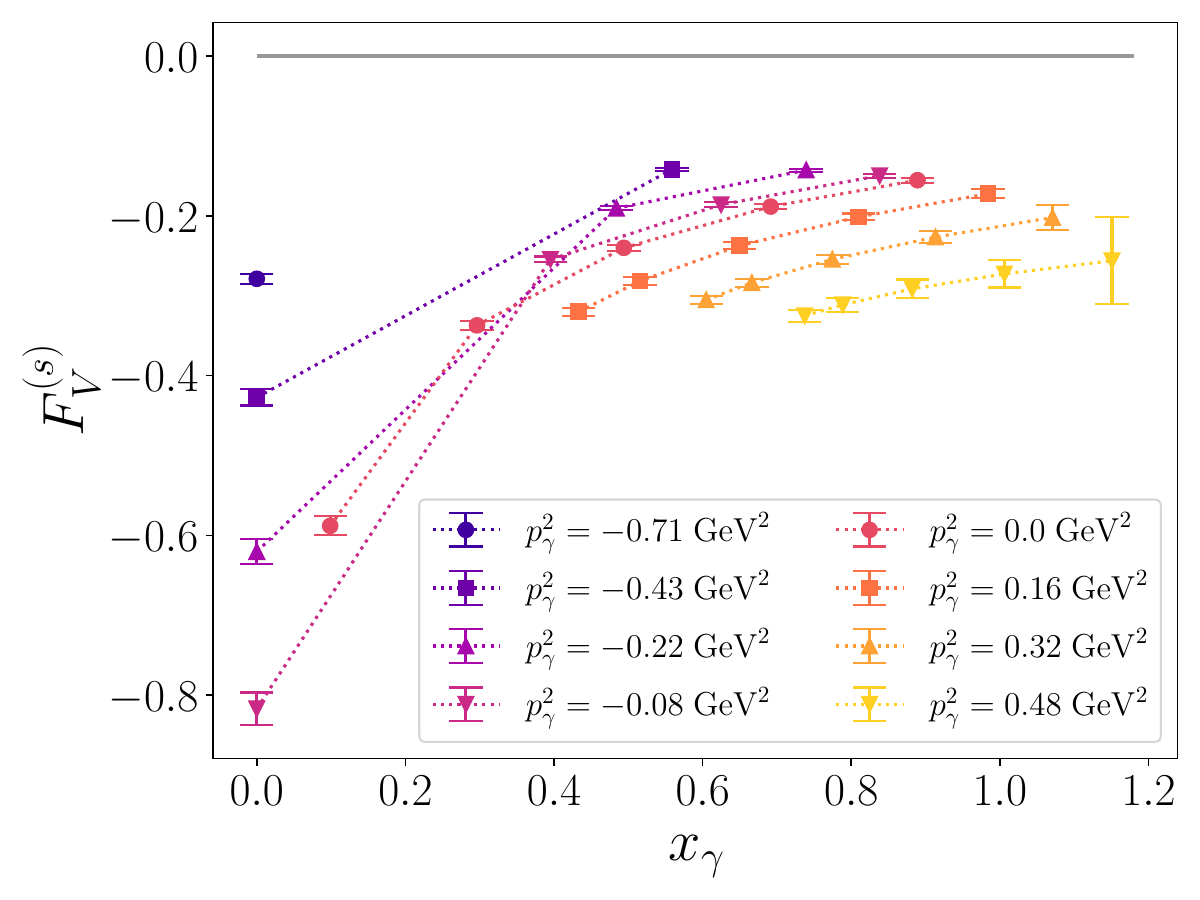}
    \includegraphics[width=0.48\linewidth]{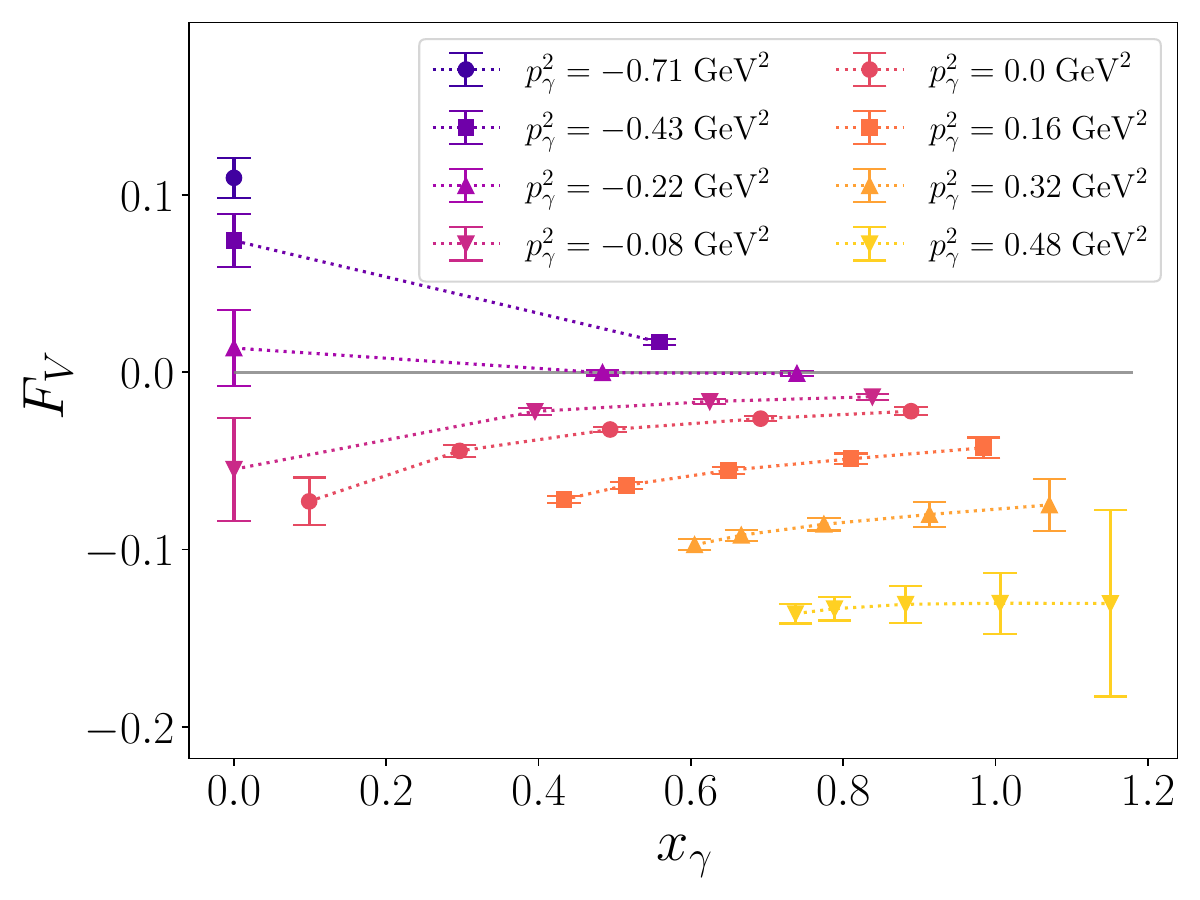}
    \caption{Different components of the vector form factor as a function of $x_\gamma$. Data with different colors and markers correspond to different values of fixed photon virtuality $p_\gamma^2$. 
    The top left and right plots show the charm ($F_V^{(c)}$) and strange ($F_V^{(s)}$) quark components of the form factor, respectively.
    The lower plot shows the full form factor $F_V = F_V^{(s)}+F_V^{(c)}$. We observe strong cancellations between the individual quark components, with almost perfect cancellation for photon virtuality $p_\gamma^2 = -0.22$ GeV$^2$.
    Note that the dashed lines connecting different points are meant only to guide the eye. The form-factor values and covariance matrices are also provided as machine-readable files in the Supplemental Material \cite{supplemental}.}
    \label{fig:FV_c_s_tot}
\end{figure}

The components of the form factor we calculate directly from our fits are the individual time orderings of the different quark current components $F_V^{(s),>}, F_V^{(s),>,{\rm EM}}, F_V^{(c),>}, F_V^{(c),>,{\rm EM}}$.
The charm and strange quark components are calculated by summing the individual time orderings $F_V^{(q)}=F_V^{(q),>,{\rm EM}} + F_V^{(q),>}$, where $q=c,s$, and the full form factor is $F_V = F_V^{(s)} + F_V^{(c)}$.
Looking at the 2d plots in Fig.~\ref{fig:indiv_time_orderings_FV}, we see that $F_V^{(c),>}$ and $F_V^{(c),>,{\rm EM}}$ have the same sign for all kinematic points, and there are no cancellations when summing them to determine $F_V^{(c)}$
A similar behavior is also observed for the strange quark component.
The same is not true, however, when adding the individual quark components $F_V^{(c)}$ and $F_V^{(s)}$ to calculate the full form factor $F_V$ (and also when adding the charm and strange quark components for individual time orderings, seen in Fig.~\ref{fig:indiv_time_orderings_FV}).
Looking at Fig.~\ref{fig:FV_c_s_tot}, we see that, for photon virtuality $p_\gamma^2 = -0.22$ GeV$^2$, there is almost perfect cancellation, with the form factor results entirely consistent with zero for all $x_\gamma$ values we studied.
Similar cancellations were observed in Refs.~\cite{Giusti:2023pot, Frezzotti:2023ygt} for $F_V$ at zero virtuality and in Refs.~\cite{Donald:2013sra,Pullin:2021ebn} for the $D_s D_s^*\gamma$ couplings.
The latter are associated with the pole residues appearing in the form factors for the $D_s \to \ell\bar{\nu}\gamma$ decay.
This cancellation becomes weaker as the magnitude of the photon virtuality is increased.
For increasing positive virtuality, we observe an enhancement in the magnitude of $F_V$. 
This is likely due to enhanced contributions from the $\phi$ meson intermediate state contributing to $F_V^{(s),>}$, which is further supported by the fact that $F_V^{(s),>}$ is also enhanced for increasing positive virtuality.

\begin{figure}
    \includegraphics[width=0.49\linewidth]{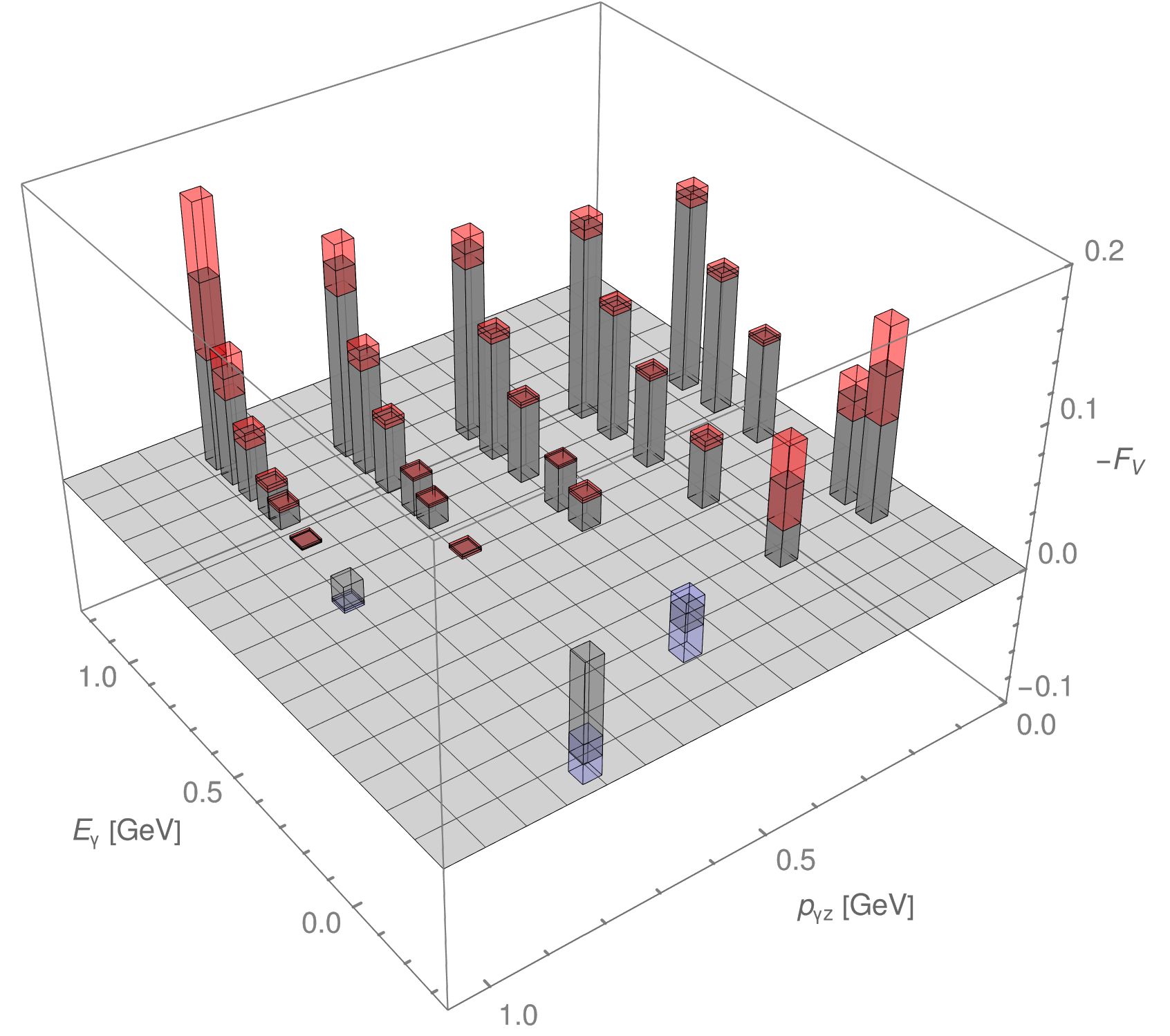} \hfill  \includegraphics[width=0.49\linewidth]{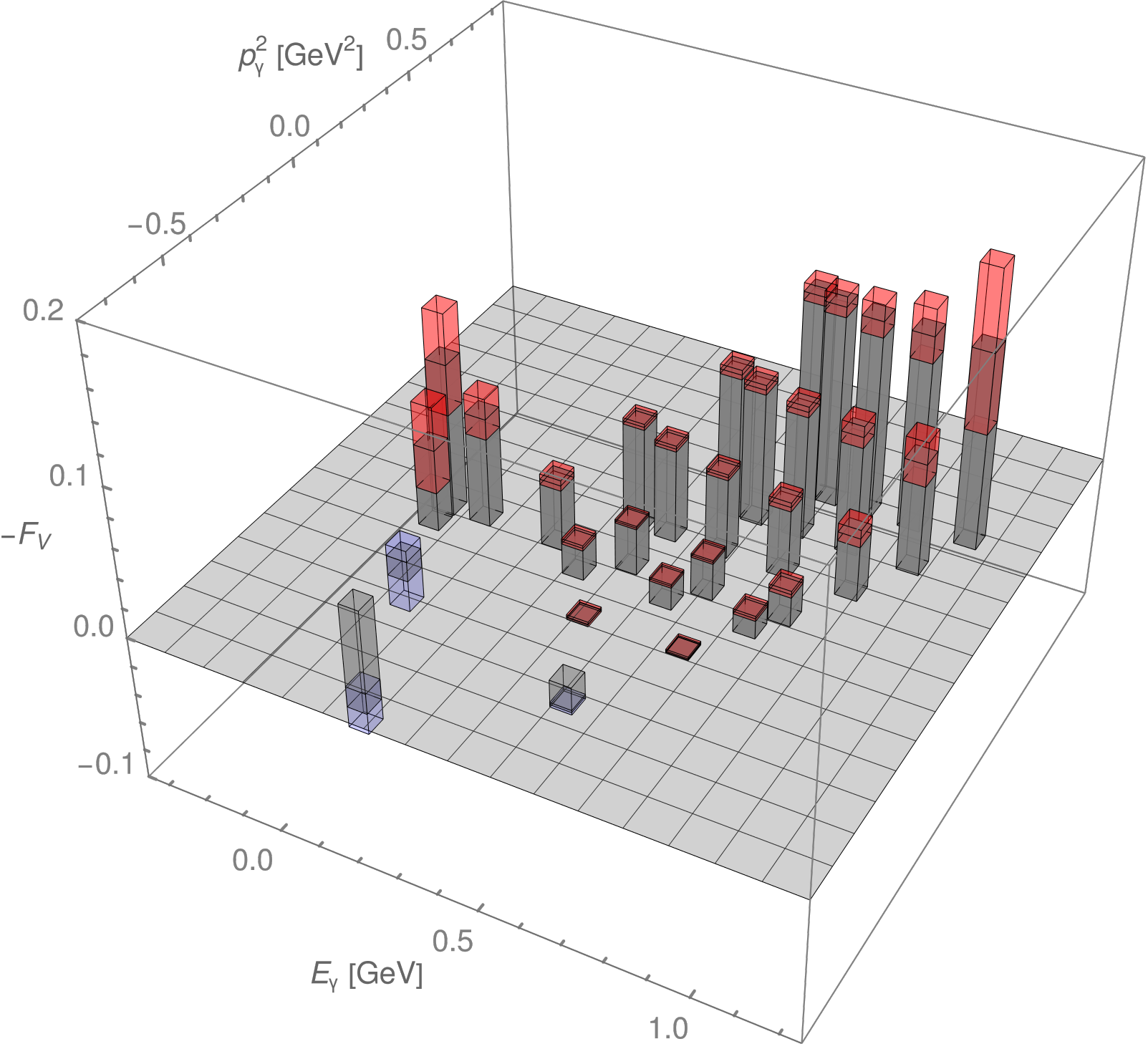} 
    \caption{The form factor $-F_V$ plotted versus $E_\gamma$ and $ p_{\gamma z}$ (left) or versus $E_\gamma$ and virtuality $ p_{\gamma}^2$ (right).  The gray bars indicate the central values, while the colored bars are the $\pm1\sigma$ statistical error bars. The color of the error bars is used to indicate the sign of the central value: red: positive and blue: negative.}
    \label{fig:3dtot}
\end{figure}

\begin{figure}
      \includegraphics[width=0.49\linewidth]{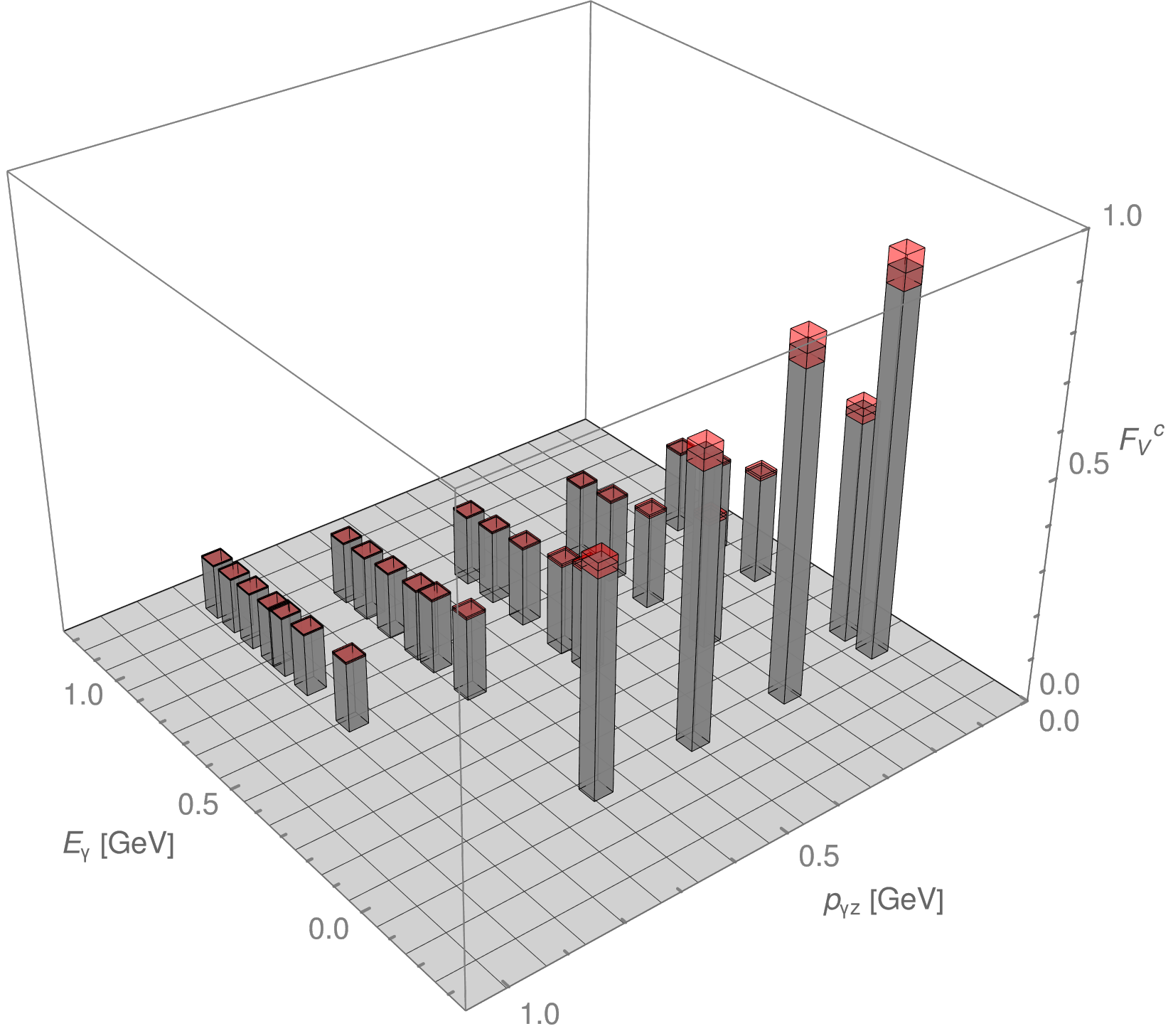}  \hfill  \includegraphics[width=0.49\linewidth]{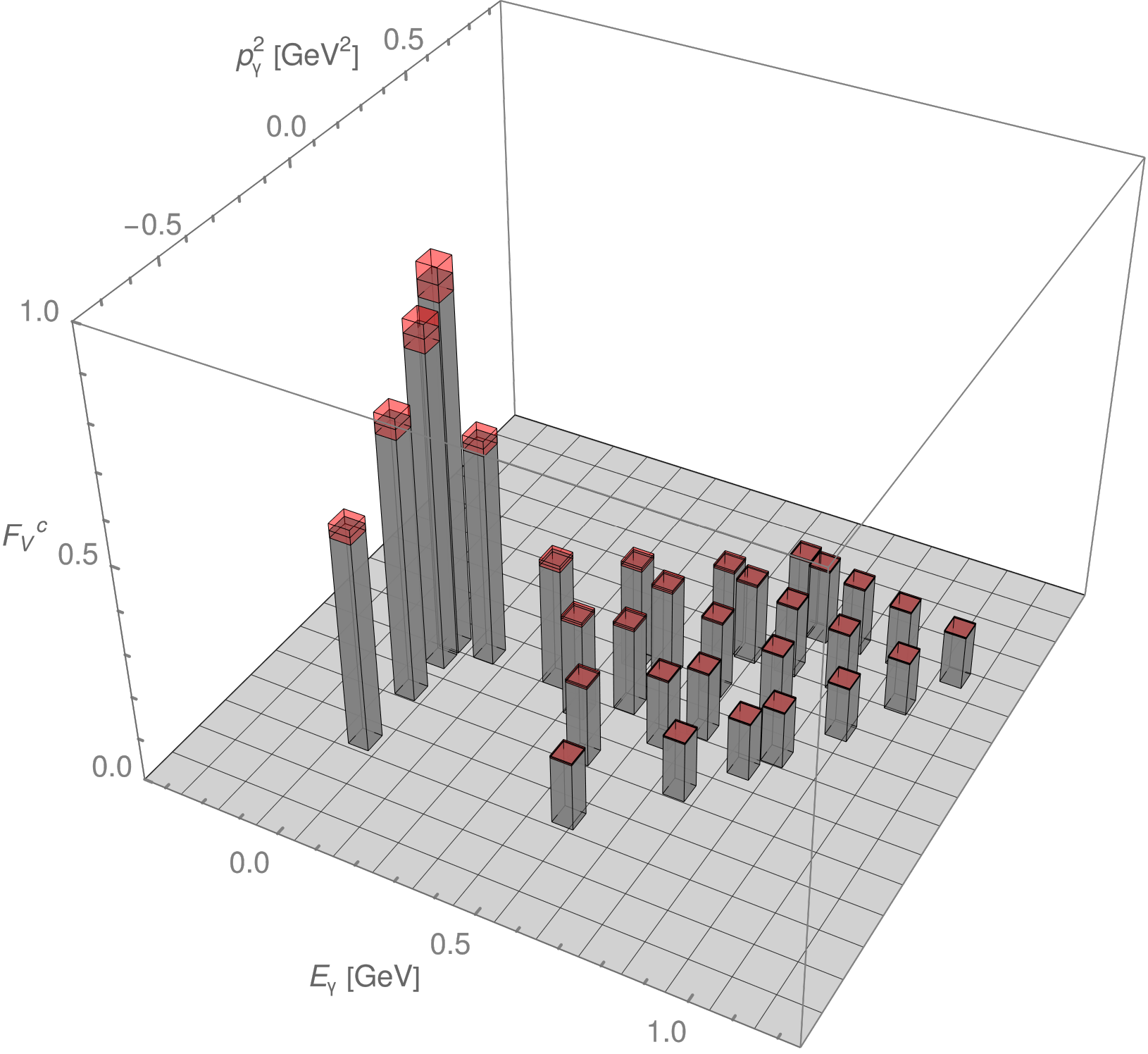} 
    \caption{Like Fig.~\protect\ref{fig:3dtot}, but for the form factor $F_V^c$.}
    \label{fig:3db}
\end{figure}

\begin{figure}
    \includegraphics[width=0.49\linewidth]{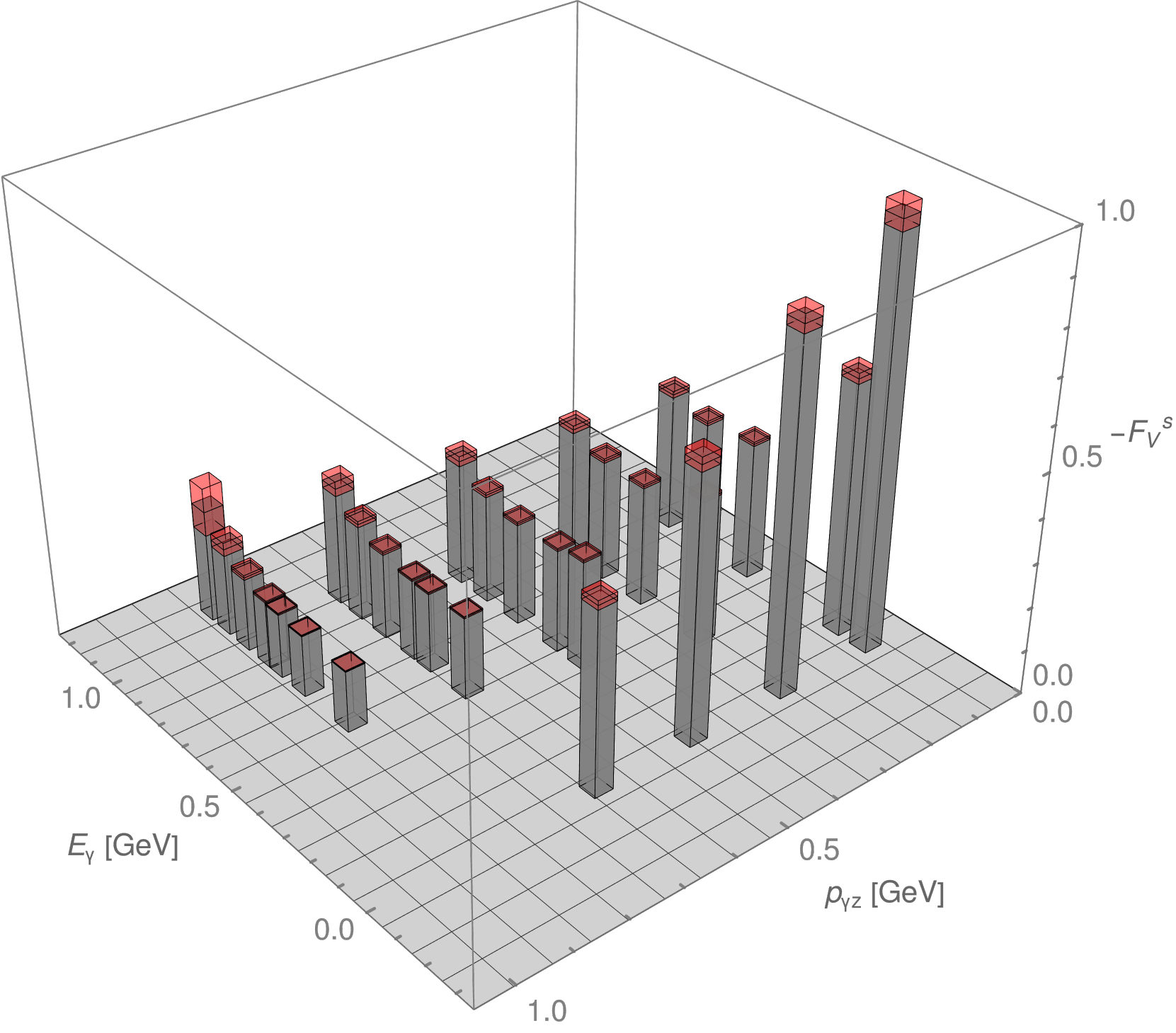}   \hfill  \includegraphics[width=0.49\linewidth]{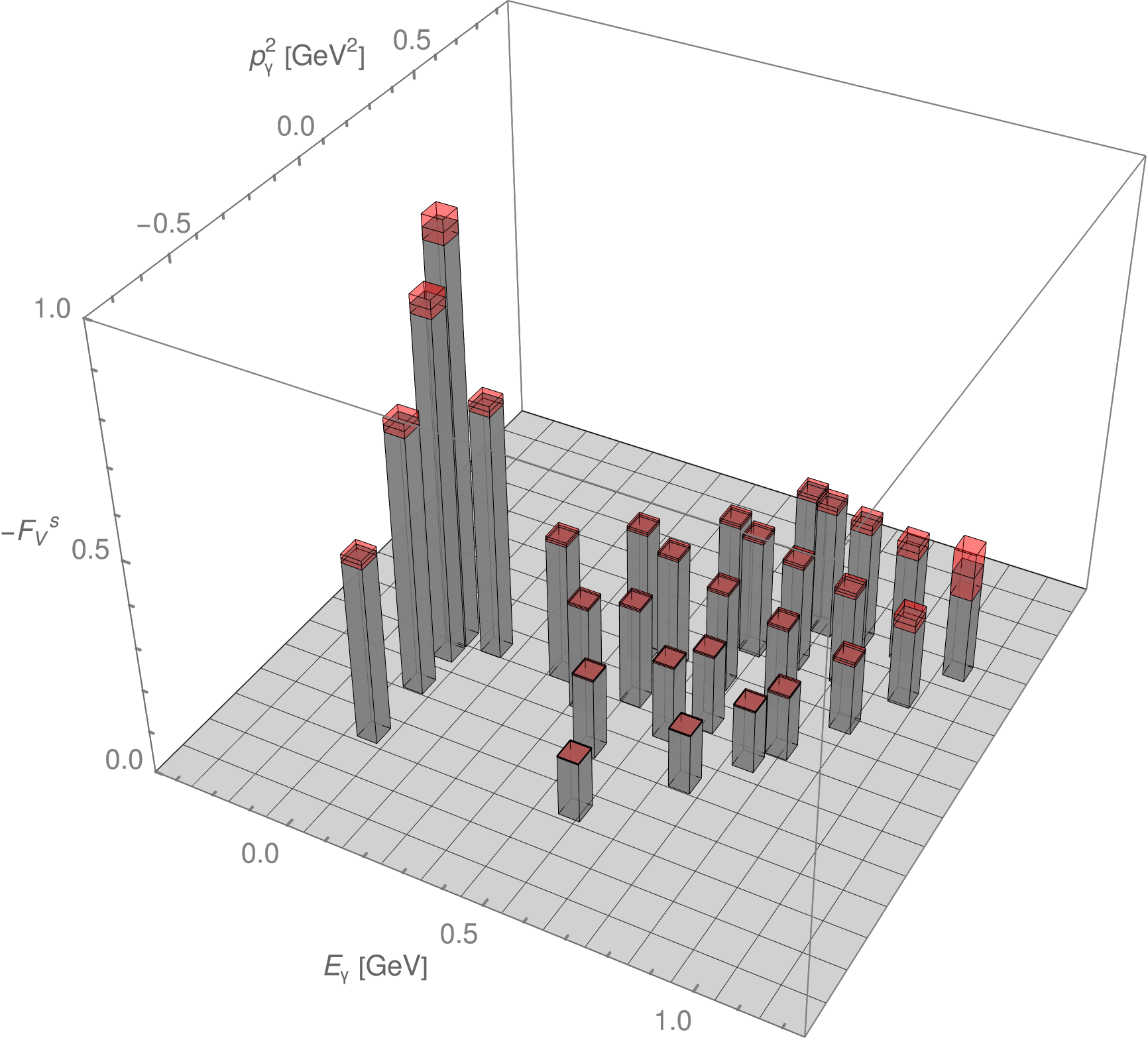} 
    \caption{Like Fig.~\protect\ref{fig:3dtot}, but for the form factor $-F_V^s$.}
    \label{fig:3du}
\end{figure}

We now discuss the form factor results in the 3d plots.
Looking at the left plot in Fig.~\ref{fig:3db}, we see that, at fixed $p_{\gamma,z}$, the charm component $F_V^{(c)}$ decreases monotonically with increasing $E_\gamma$.
As seen in the left plot in Fig.~\ref{fig:3du}, the same is not true for the strange quark component $F_V^{(s)}$, which exhibits non-monotonic behavior for fixed $p_{\gamma,z}$ and increasing $E_\gamma$.
Turning now to the full form factor $F_V$ in Fig.~\ref{fig:3dtot}, first we observe a significant increase in the relative size of the uncertainties, which results from the strong cancellations between the charm and strange quark components.
Furthermore, the behavior of $F_V$ is generally more interesting than the individual quark components.
In particular, for small fixed momentum, the vector form factor starts positive, decreases, and then increases as $E_\gamma$ is increased.
In contrast, for large photon momentum, $F_V$ starts out negative, crosses zero, and continues to increase.
At fixed virtuality $p_\gamma^2$, the form factor generally decreases in magnitude with increasing photon energy $E_\gamma$.

\FloatBarrier
\section{Conclusions}
\label{sec:conclusions}

In this extension of the work in Ref.~\cite{Giusti:2023pot}, we have demonstrated that, using the 3d method, lattice QCD can be used to calculate three-point functions for radiative-leptonic decays into virtual photons for zero additional propagator solves relative to the real photon case.
Another important component of our work is the study of not only positive photon virtualities, which are relevant for $H^+ \to \ell'^+ \ell'^- \ell^+ \nu_\ell$ and $H^0 \to \ell^{\prime +} \ell^{\prime -} \ell^+ \ell^-$ decays, but also \emph{negative} virtualities, which can provide valuable input for controlling soft contributions to heavy-meson form factors when using QCD factorization approaches.
Interestingly, unlike the positive virtuality case which has fundamental limitations on the values of virtuality that can be studied using lattice QCD ($p_\gamma^2 < 4m_\pi^2$), one can study negative virtualities up to the theoretically smallest value associated with $E_\gamma=0$.
With our methods, using only $N_{\rm cfg}=25$ gauge field configurations, we obtain precise values of $F_V$ over most of the kinematic ranges studied.

To demonstrate our methods, we calculated the vector form factor $F_V$ for a range of photon momenta and virtualities.
We provided a detailed discussion of the behavior of the unwanted exponentials that come with each intermediate state in the spectral decomposition, focusing on their behavior in different kinematic regimes.
We showed that, for each time ordering, there are extreme kinematic regimes where the unwanted exponentials decay slowly and become difficult to remove through extrapolations. 
This was particularly problematic for the $t_{\rm em}<0$ time ordering where one integrates towards the interpolating field.
To overcome this difficulty, we calculated an alternative three-point function, where, instead of the weak current being fixed to the origin, the EM current is fixed to the origin.
Using this alternative three-point function, we were able to reliably control the unwanted exponentials associated with both time orderings, as shown by performing a detailed stability analysis of the fit result as a function of the fit ranges used.
An interesting followup research direction is the inclusion of the $t_{\rm em}<0$ and $t_W<0$ time orderings in the analysis; while doing so will require careful treatment to ensure the unwanted exponentials are appropriately controlled, it could result in some improvement in the statistical precision achieved.

This work lays the groundwork for future calculations of radiative-leptonic decays into virtual photons.
The next step is to determine the 3 axial form factors, which was done for kaon radiative leptonic decays at positive virtuality in Ref.~\cite{Gagliardi:2022szw}.
While this step may appear straightforward, it is complicated by the fact that one must develop methods to remove lattice artifacts that diverge as $\mathcal{O}(a^n/E_\gamma)$ when isolating the structure-dependent component of these form factors.
An initial method for doing so was proposed in Ref.~\cite{Gagliardi:2022szw}. 
Another important step is the calculation of the disconnected diagrams, whose contribution will likely become enhanced as one approaches $p_\gamma^2 \sim 4 m_\pi^2$.
With these developments in hand, the next step is to perform calculations on several ensembles and extrapolate to the continuum and physical pion mass, which will provide precise Standard-Model predictions. Lattice QCD is perhaps the only reliable approach for charm radiative-leptonic decays, as neither chiral perturbation theory nor heavy-quark effective  theory are expected to converge well in this sector.
The ultimate goal of this program is to study decays of $B$ and $B_s$ mesons and compare with QCD-factorization/effective-field-theory studies.
With the advent of ensembles with extremely fine lattice spacings~\cite{Aoki:2025etz}, our methods can be directly applied by performing the calculation for a variety of heavy quark masses and extrapolating to the physical bottom quark mass.

\section*{Acknowledgements}

We thank the RBC and UKQCD Collaborations for providing the gauge-field configurations.  We are grateful to Martin Beneke and Vladimir Braun for valuable discussions related to this work.
C.F.K.~is supported in part by the Department of Physics, Maryland Center for Fundamental Physics, and the College of Computer, Mathematical, and Natural Sciences at the University of Maryland, College Park, and was supported by the U.S. Department of Energy, Office of Science, Office of Advanced Scientific Computing Research, Department of
Energy Computational Science Graduate Fellowship under Award Number DE-SC0020347. 
S.M.~is supported by the U.S.~Department of Energy, Office of Science, Office of High Energy Physics under Award Number DE-SC0009913.
A.S.~was supported in part by the U.S.~DOE contract \#DE-SC0012704. 
This research used resources provided by the National Energy Research Scientific Computing Center, a DOE Office of Science User Facility supported by the Office of Science of the U.S. Department of Energy under Contract No.~DE-AC02-05CH11231. This research also used resources at Purdue University RCAC and at the University of Texas TACC through the Extreme Science and Engineering Discovery Environment (XSEDE) \cite{XSEDE} and the Advanced Cyberinfrastructure Coordination Ecosystem: Services \& Support (ACCESS) program \cite{10.1145/3569951.3597559}, which are supported by U.S. National Science Foundation grants ACI-154856, 2138259, 2138286, 2138307, 2137603, and 2138296. We acknowledge PRACE for awarding us access to SuperMUC-NG at GCS@LRZ, Germany.

\bibliography{main}{}
\bibliographystyle{utphys-noitalics}

\end{document}